\documentclass[journal]{IEEEtran}
\UseRawInputEncoding
\usepackage{cite}
\usepackage{amsmath,amssymb,amsfonts}
\usepackage{graphicx}
\usepackage{textcomp}
\usepackage{xcolor}
\usepackage{soul}
\usepackage{subcaption}
\usepackage{multirow}
\usepackage{makecell}
\usepackage{booktabs} 
\usepackage[usestackEOL]{stackengine}
\usepackage{amsmath}
\usepackage[linesnumbered,ruled]{algorithm2e}
\usepackage{algpseudocode}
\usepackage[euler]{textgreek}
\usepackage{svg}
\usepackage{hyperref}
\usepackage[symbol]{footmisc}
\def\BibTeX{{\rm B\kern-.05em{\sc i\kern-.025em b}\kern-.08em
  T\kern-.1667em\lower.7ex\hbox{E}\kern-.125emX}}
\begin{document}

\newcommand{\blue}[1]{{{#1}}} 
\newcommand{\red}[1]{{{\color{red} #1}}}
\newcommand{\magenta}[1]{{{\color{magenta} #1}}}
\newcommand{\violet}[1]{{{\color{violet} #1}}}
\newcommand{\hyoseung}[1]{{\blue{\textbf{(Hyoseung: #1)}}}}
\newcommand{\green}[1]{{{#1}}} 

\title{OpenSense: An Open-World Sensing Framework for Incremental Learning and Dynamic Sensor Scheduling on Embedded Edge Devices}


\author{
    Abdulrahman~Bukhari,~\IEEEmembership{Student Member,~IEEE,}
    Seyedmehdi~Hosseinimotlagh, \\
    Hyoseung~Kim,~\IEEEmembership{Member,~IEEE}
    \thanks{This work is supported by grants from NSF 1943265, NIJ 2019-NE-BX-0006, USDA/NIFA 2020-51181-32198.}    
    \thanks{A. Bukhari and H. Kim are with the Department of Electrical and Computer Engineering, the University of California Riverside, Riverside, CA 92521, USA (email: \href{mailto:abukh001@ucr.edu}{abukh001@ucr.edu}; \href{mailto:hyoseung@ucr.edu}{hyoseung@ucr.edu}). S. Hosseinimotlagh is currently with eBay. The work was done while he was a postdoc at the University of California Riverside (email: \href{mailto:shoss007@ucr.edu}{shoss007@ucr.edu}). 
    }

}

\maketitle

\begin{abstract}
\blue{Recent advances in Internet-of-Things (IoT) technologies have sparked significant interest towards developing learning-based sensing applications on embedded edge devices. These efforts, however, are being challenged by the complexities of adapting to unforeseen conditions in an open-world environment, mainly due to the intensive computational and energy demands exceeding the capabilities of edge devices.
In this paper, we propose OpenSense, an open-world time-series sensing framework for making inferences from time-series sensor data and achieving incremental learning on an embedded edge device with limited resources. The proposed framework is able to achieve two essential tasks, inference and incremental learning, eliminating the necessity for powerful cloud servers. 
In addition, to secure enough time for incremental learning and reduce energy consumption, we need to schedule sensing activities without missing any events in the environment. 
Therefore, we propose two dynamic sensor scheduling techniques: (i) a class-level period assignment scheduler that finds an appropriate sensing period for each inferred class, and (ii) a Q-learning-based scheduler that dynamically determines the sensing interval for each classification moment by learning the patterns of event classes.
With this framework, we discuss the design choices made to ensure satisfactory learning performance and efficient resource usage. Experimental results demonstrate the ability of the system to incrementally adapt to unforeseen conditions and to efficiently schedule to run on a resource-constrained device.}
\end{abstract}


\maketitle
\pagestyle{plain}

\section{Introduction}
\blue{As technology continues to evolve, sensing is becoming an integral part of environmental and personal applications. Enhanced computing capabilities combined with modern machine learning techniques enable Internet-of-Things (IoT) systems to achieve smart sensing. This allows them to process data, from its acquisition to inference, providing nearly real-time insights to the end-users. For instance, health trackers, using instruments like accelerometers and gyroscopes within smartphones and wearable devices, monitor everyday activities including step counts and different postures. However, a notable challenge remains: while adept at providing accurate results for previously-seen data, e.g., input data belonging to one of the classes of training data, these systems often necessitate server-based interventions to update their learned models when encountered with novel, unforeseen data. }

In conventional smart sensing applications,  raw sensor data is transferred to an edge device for additional processing. Some applications require learning algorithms to grasp a higher level of interaction between the sensor data and the application requirements. 
For example, Synthetic Sensors~\cite{laput2017synthetic} collect and process raw data on an embedded device before sending it to the server for SVM-based learning and inference; 
DeepSense~\cite{DeepSense} can be applied to accelerometer and gyroscope data to learn and recognize human activities through a deep neural network (DNN). If these sensing frameworks encounter unlabeled data that belong to a new, unseen class by the learning model, it will be incorrectly classified as one of the existing classes. 
Finding new classes in the environment and incorporating them into the model has traditionally been performed by manually labeling new data and retraining the whole model using both old and new labeled data.


Updating a model over time is an active research problem, called incremental learning, and it has been widely studied especially for vision applications. Although incremental learning algorithms such as iCaRL~\cite{Rebuffi2017iCaRLIC} can add new classes and update network parameters for them, those new classes must be identified by the user and unseen data must be labeled accordingly before each model update. Thus, in this paper, we specifically call them {\em supervised incremental learning}. These limitations introduce an open-world learning problem~\cite{Jafarzadeh2020OpenWorldLW}, where the system has to differentiate by itself whether new data belong to an unknown class or a known class and update the model accordingly without sacrificing the inference performance. Since open-world learning methods do not require any human intervention, we call them {\em unsupervised incremental learning}. 
Recent work in both supervised and unsupervised incremental learning has shown significant advantages over offline pretrained models.  However, most of these approaches are computationally hungry and require powerful machines for training. Additionally, they depend on the base machine-learning model, e.g., a DNN, which can be often too heavy to use on edge devices.

\blue{
There is another challenge in edge-based smart sensing frameworks.
Consider a system with incremental learning capabilities for smart homes. The system consists of a set of sensors and one edge device, with each sensor communicating with the edge device that performs event classification and incremental learning.
The edge device has to determine when and how often to pull sensor data from the sensors, which we call {\em sensor scheduling}.
Since the edge device can classify the current event in the environment only when the updated sensor data arrives, a shorter sensing interval is better for classification performance. On the other hand, a longer interval is better to reduce energy for data transmission and to utilize the resulting idle time for incremental learning which is essential to detect new event classes.
While many sensing frameworks~\cite{laput2017synthetic,DeepSense} assume the collection of sensor data at fixed time intervals, e.g., every second, the use of fixed sensing intervals not only negatively affects classification performance and energy usage, but also introduces a significant overhead to incremental learning tasks, making them unable to update the model at runtime. 
}

To enable a sensing framework to incrementally learn new classes in both supervised and an unsupervised fashion while addressing the aforementioned challenges, this paper proposes OpenSense, an open-world sensing framework for resource-constrained embedded edge devices. OpenSense can identify (thus reject from inference) unknown samples, assign new classes to these rejected samples, incorporate the new classes, and update the model incrementally on an edge device. \blue{In addition, we propose dynamic sensor schedulers to determine appropriate sensing intervals such that the edge device can produce timely classification results, reduce energy consumption, and preserve time for incremental learning tasks to run.} The contributions of this work are as follows:
\begin{itemize}
\item We present the OpenSense framework that can run incremental learning algorithms for both supervised and unsupervised time-series sensing data problems.

\item We propose an efficient DNN architecture called sDNN, which outperforms the state-of-art architecture in both inference performance and resource efficiency for time-series activity classification.

\blue{
\item We propose two novel sensor scheduling algorithms, class-level period assignment and Q-learning-based scheduler, to reduce the energy consumption of the edge device while increasing the idle time on the edge device to perform learning tasks without sacrificing the timeliness of classification.}

\item We demonstrate the implementation of OpenSense on a resource-constrained edge device and its effectiveness in open-world incremental learning of time-series data.

\end{itemize}

\blue{This is an extend version of our prior work~\cite{bukhari2022} with the following new contributions: (i) we propose a reinforcement learning-based dynamic sensor scheduler that aims to solve the issues of static periods assigned by the class-level period assignment scheduler, (ii) we present additional experimental results to compare the performance of the two proposed sensor schedulers, and (iii) we analyze the overhead imposed by each scheduler on a representative edge device.}

\section{Related Work}

\subsection{Supervised Incremental Learning}
Incremental learning techniques have to overcome two challenges to maintain their performance. The first challenge is to learn new classes from a stream of data without ``catastrophic forgetting''~\cite{MCCLOSKEY1989109}, in which the new data causes the neural network to forget what has been learned from the previous data. Early attempts made by~\cite{xiao2014error-driven,rusu2016pnn} tackle this challenge by expanding the neural network model as more classes are learned. However, these approaches introduce a new challenge of ever-increasing computation and memory demands as the model will keep growing incrementally, which is particularly problematic for embedded systems.

A number of approaches have been proposed to overcome the aforementioned problems of incremental learning. Few-Shots Class-Incremental Learning (FSCIL)~\cite{zhang2021few} aims to learn both new and old classes with a limited number of training data, which limits the capacity of the model while avoiding forgetting observed classes in a stream of data. iCaRL~\cite{Rebuffi2017iCaRLIC} can learn new classes with a fixed feature representation based on the nearest mean classifier algorithm and the distillation loss to maintain the performance without forgetting. Due to the nature of these approaches, they perform well only under the assumption that the data set used for training is labeled by the user for both old and new classes.

\subsection{Unsupervised Incremental Learning}
Contrary to the supervised approaches, classifiers for unsupervised incremental learning (a.k.a. open-world learning) must be able to distinguish unknown (unlabeled) samples in the input data stream and classify them as new classes, while maintaining the performance of classifying old classes. \green{Recently, researchers have shown more interest in the open-world problem~\cite{Joseph_2021_CVPR, ismail2023enhancing,zhao2023revisiting,gutoski2023unsupervised}}. Early works~\cite{Hassen2020LearningAN,Bendale2016TowardsOS} have established a stepping stone toward this. In~\cite{Bendale2016TowardsOS}, the authors introduce the open-set problem for DNNs and propose a layer called OpenMax that extends the softmax layer to find the likelihood that an input sample is of an unknown class. A more sophisticated scheme called the Extreme Value Machine (EVM) is proposed by Ethan et al.~\cite{Rudd_2018} based on the Extreme Value Theorem~\cite{kotz2000extreme}. EVM can fit an initial set of data into extreme vectors (EVs) using the Weibull distribution. Once EVM is trained, it can be updated incrementally by obtaining new EVs. In addition to its inference performance in open-world settings, the major advantage of EVM is the ability to limit the size of the model by setting the number of EVs in the model.

In~\cite{Jafarzadeh2020OpenWorldLW}, the authors extend the concept of EVM and re-define the evaluation protocol for open-world learning. They claim that an open-world learner must recognize both known and unknown classes and classify the unknown classes into new classes without forgetting classes previously learned. They also proposed a new metric, the Open-World Metric (OWM), since the testing set should include both known and unknown classes, thereby requiring a novel measurement that captures both the accuracy of the known classes and the classification of unknown classes. The authors in~\cite{dhamija2021ssfowl} introduce the notion of self-supervised features for open-world learning in order to avoid the overlap between known and unknown classes in the supervised feature space. Another work by Joseph et al.~\cite{9578456} extends the open-world learning to the problem of identifying an unknown object in images, by using an energy-based model~\cite{LeCun06atutorial} to create a separation between known and unknown objects in the energy space.

Although these approaches have shown acceptable performance in image classification, to the extent of our knowledge, none has been applied to time-series multisensor IoT sensing applications which this paper focuses on.

\subsection{Time-Series Sensing Frameworks}
Extending supervised or unsupervised incremental learning to sensing frameworks requires a state-of-the-art classifier that on one hand has a high inference performance across different applications, and on the other hand is deployable on embedded edge devices such as Raspberry Pi. There are mainly two approaches for sensing applications: special-purpose~\cite{Klingensmith2014hotcold,Kuznetsov2010upstream,scott2011preheat,Mohsen-iotj,
karimi2020energy,IENCO202336} and general-purpose sensing~\cite{1677514,5993550,laput2017synthetic,Gadaleta2018IDNetSG,Zahin2019SensorBasedHA}. 

The special-purpose sensing approach uses a single sensor to measure one aspect of the environment, e.g., temperature. Setting a threshold in these sensors is usually sufficient for controlling or monitoring applications and does not require complex learning-based techniques. The general-purpose sensing approach uses a combination of two or more sensors to recognize multiple events or activities in the scene, often through machine learning. Laput et al.~\cite{laput2017synthetic} propose a custom sensor tag including 9 sensors to detect 38 different events, such as kettle on/off, door open/close, phone ringing, etc., using a simple SVM trained for each event. 
DeepSense~\cite{DeepSense} is a DNN-based time-series mobile sensing framework that can be applied to regression and classification problems by extracting both spatial and temporal features/relationships using CNN and RNN. DeepSense outperforms other traditional learning methods in different classification problems such as human activity recognition and user identification. However, it requires longer training time due to the complexity of the architecture and does not support incremental learning in open-world settings. We address these limitations in this paper.

\subsection{Sensor Scheduling}
\blue{Scheduling to reduce energy consumption~\cite{Heo2017rtifttt,heo2020sharingaware,Rashtian2020sdcps,Bukhari23} or to meet tasks deadlines~\cite{Mohsen-iotj,karimi2020energy,mohsen2022} has been broadly discussed in sensing and IoT frameworks. Heo at al.~\cite{heo2020sharingaware} used data freshness as a constraint for acquiring sensor data and preventing the redundancy of data from shared sensors. The same authors also proposed a sensor polling scheduler based on the rate of changes in sensor values and the probability of whether a trigger condition will occur soon or not~\cite{Heo2017rtifttt}. However, it primarily focuses on slowly-changing environmental sensors, e.g., temperature, humidity and UV index, and cannot be used for high-level inference queries beyond raw-sensor trigger conditions. Rashitan et al.~\cite{Rashtian2020sdcps} proposed to assign priority to each sensor in a sensor network to reduce the amount of data transmitted to a server and workload intensity. However, this work does not consider the impact of data inference and model updates. 

Studies in \cite{Mohsen-iotj,karimi2020energy,mohsen2022} focus on real-time sensing task scheduling in the context of energy-harvesting batteryless devices since it is essential to meet the energy constraints for these devices. For example, Mohsen et al.~\cite{mohsen2022} proposed adjusting sensing periods of tasks to satisfy data freshness constraints based on the energy availability of the device. There are other studies that used reinforcement learning for sensing systems~\cite{frat02,AitAoudia2018RLManAE}. However, their focus is mainly on learning the energy supply patterns of the device, not on sensor scheduling and event classification. In this paper, we aim to develop dynamic sensor scheduling algorithms that can efficiently collect sensor data to detect and classify events with random occurrence rates. }

\section{OpenSense Framework}

\begin{figure}[t]
\centerline{\includegraphics[width=0.5\textwidth]{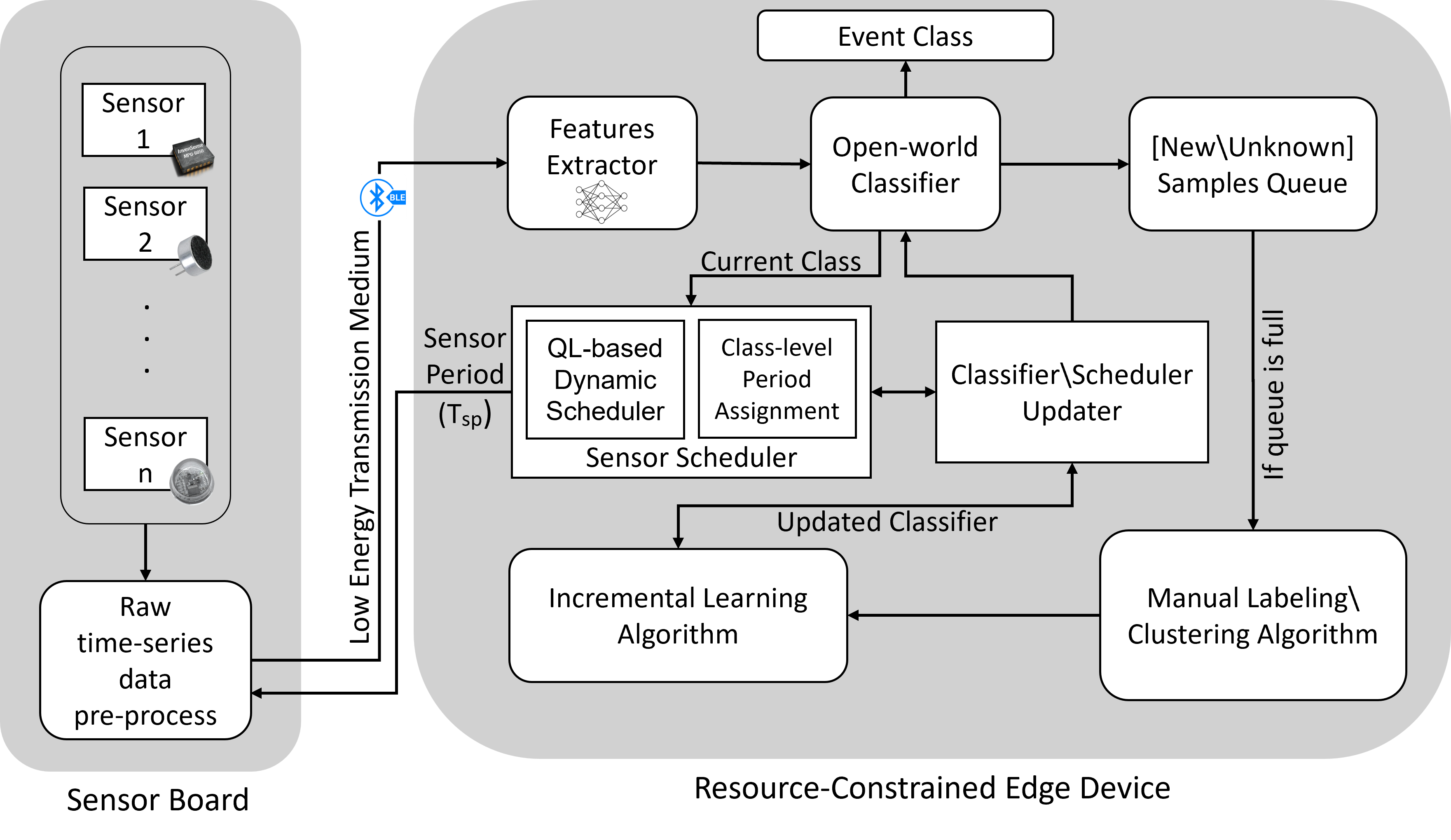}}
\caption{\green{Block diagram of the proposed OpenSense framework}}
\label{fig2}
\end{figure}

The main objectives of the proposed OpenSense framework\footnote{The source code is available at: \url{https://github.com/rtenlab/OpenSense}.} are to characterize raw time-series sensor data into informative labels to the user and to incrementally learn new classes that have not been seen by the classifier in both supervised and unsupervised manners. The framework consists of two units: a sensor board attached to the point of interest to acquire data from the application environment, and a resource-constraint edge device that is responsible for performing inference and incrementally learning novel classes as shown in Fig.~\ref{fig2}. Unlike traditional time-series sensing frameworks, this framework does not require a cloud server to learn new classes and update the model. \green{Except for the initial DNN and sensor scheduler training, which is performed on a Linux-based machine equipped with a GPU due to the need for access to the complete training dataset and the high computational demands, all procedures are running on the edge device.}


\subsection{Sensor Board and Edge Device}
\label{sec:sensor_board}

\noindent\textbf{Sensor Board:} 
The sensor board collects and preprocesses raw sensor data samples, and then transmits them periodically in a specific size of window to the edge device. There can be one or more sensor boards in the scene. Each sensor board consists of a single sensor or an array of sensors and a low energy transmission medium with a processing unit to buffer and prepare the raw time-series data. The number and type of sensors in the board is decided based on application requirements. Multiple different sensors are used in most recent applications; for example, a combination of accelerometer, microphone, and illumination sensors enable capturing events in the kitchen, such as whether the faucet, kettle, microwave, air vent are running or turned off.

The processing unit prepares raw data based on the frequency of the corresponding sensor to reduce noise and overhead on data transmission without impacting the overall framework performance. A simple moving average window has shown reliable performance for sensors with high sampling frequencies, e.g., accelerometer. After prepossessing the sensors data, they are stacked together and transmitted periodically to the edge device through a low-energy data transmission medium, such as Bluetooth Low Energy (BLE). The transmission period between the sensor board and the edge device is set by the sensor scheduler (Sec.~\ref{sec:dynamic_scheduler}). 

\smallskip
\noindent\textbf{Edge Device:} 
The resource-constraint edge device is the heart of the framework. When sensor data samples are received, the edge device processes them for {\em feature extraction} using a DNN trained network (Sec.~\ref{sec:feature_extractor}). The {\em open-world classifier} (Sec.~\ref{sec:open_world_classifier}) then uses these features to distinguish whether a sample belongs to an existing class provided by the pretrained dataset or it is a new or an unknown class that has not been seen the classifier. Such samples are collected in a queue to be either manually labeled (supervised) or clustered into different classes (unsupervided) for the classifier to incrementally learn them through an incremental learning algorithm (Sec.\ref{sec:incremental_learning}). In addition, the proposed sensor schedulers (Sec.~\ref{sec:dynamic_scheduler}) and \blue{the classifier and scheduler updater (Sec.~\ref{sec:classifier_updater})} are included to change the sensor transmission period and trigger model updates, respectively.
The rest of this section describes each component in detail.


\subsection{Feature Extraction}
\label{sec:feature_extractor}
The data received from the sensor board requires further preprocessing before extracting features for inference and novel class detection. Time-series data from different sensors are processed based on the sampling rate of the corresponding sensor. Fast-Fourier Transform (FFT) is applied to data acquired from high-sampling sensors to capture the frequency patterns regardless of time dependency~\cite{DeepSense}. Statistical information, such as mean, variance, and peaks, is used for low-frequency sensors data like heart-rate and light-intensity sensors. 


\begin{figure}[tbp]
\centerline{\includegraphics[width=0.5\textwidth]{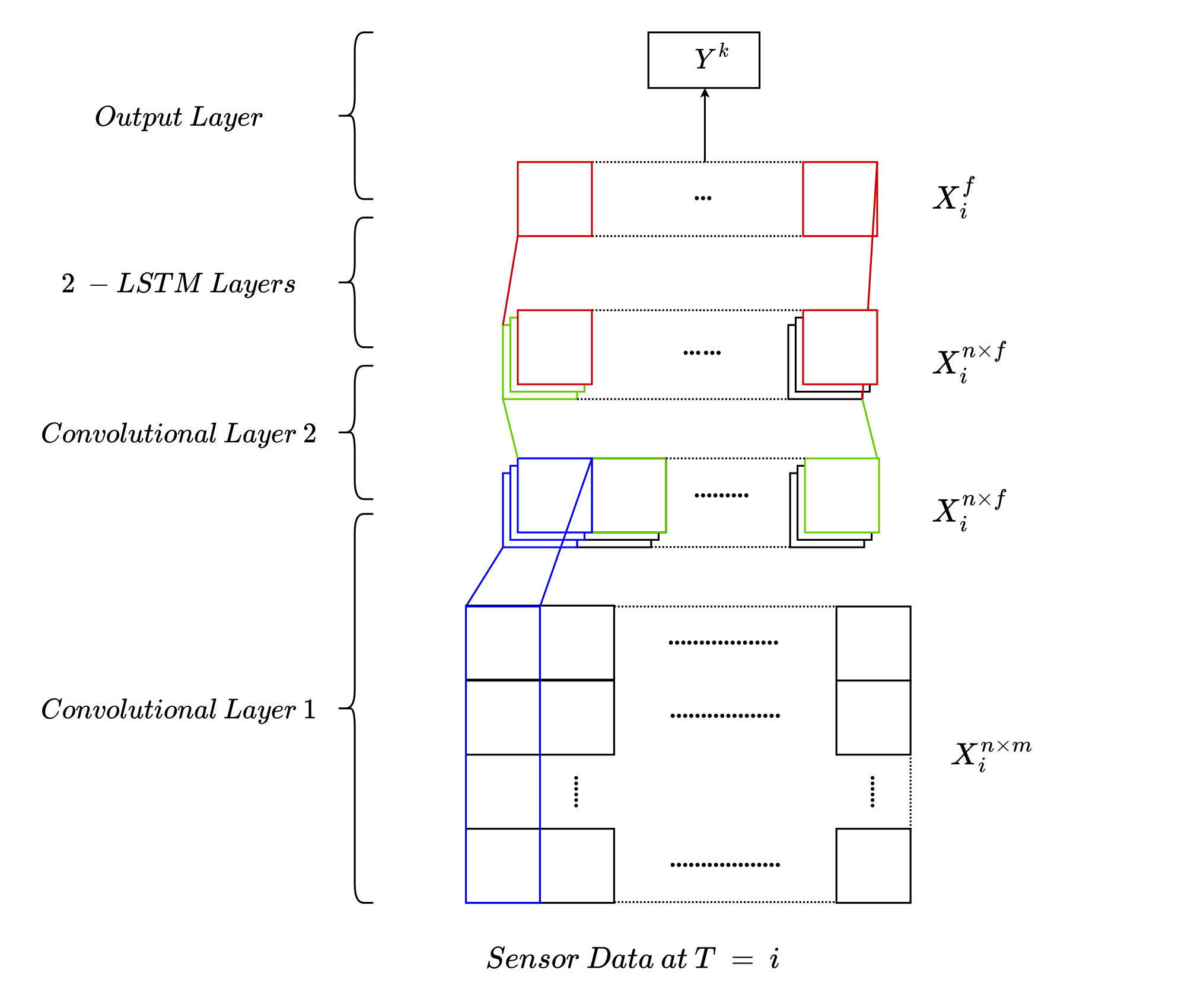}}
\vspace{-5pt}
\caption{sDNN architecture}
\label{figDNN}
\end{figure}


We perform feature extraction based on our simple DNN architecture, called sDNN, which can extract reliable features from time-series sensor data while providing runtime efficiency in inference and training. Fig.~\ref{figDNN} shows the sDNN architecture. 
The two convolutional layers first capture spatial relationships between preprocessed sensors data, and the following LSTM layers capture temporal relationships between time-series data slices. The input size of sDNN depends on the preprocessed sensor data, where $n$ is the number of time-series slices and $m$ is the length of each slice.  
Compared to other state-of-the-art architectures, such as DeepSense~\cite{DeepSense}, our proposed DNN ensures a fixed-size model and a feature vector with length $f$, where $f$ is the filter size of convolution and LSTM layers. On the other hand, the number of layers in DeepSense and the length of extracted features depend on the number of sensors on the device, making it more complex and longer as more sensors are used.
Despite its simplicity, sDNN can provide competitive and often  better inference accuracy than DeepSense as we will show in the evaluation. Note that such properties are important for incremental learning, especially on embedded edge devices.


sDNN is trained using available training datasets before deployment. 
The trained model is then used at runtime to extract features from preprocessed sensor data for the open-world classifier, by taking the output of the last layer in the DNN before the dense layer. The features extractor is also used for the initial training dataset to train the open-world classifier.

\subsection{Open-World Classifier} 
\label{sec:open_world_classifier}
The purpose of an open-world classifier is not only to recognize data samples that belong to known classes, but also to recognize unknown samples and reject them from inference. 
The open-world classifier is initially trained on the features extracted from the training datasets and deployed to the edge device. 

The built-in classifier of sDNN might be sufficient if there is no unknown sample. However, in the presence of unknown samples, it can misclassify them into one of the known classes, making it unable to be used for open-world classification. To address this issue, we adopt the Extreme Value Machine (EVM)~\cite{Rudd_2018} as the open-world classifier of our framework. 
The reasons why we chose EVM here are that (i) it can be used for both supervised and unsupervised incremental learning scenarios, (ii) it avoids the need to update the entire DNN model, mitigating the catastrophic forgetting phenomena, and (iii) it has been shown to be effective in rejecting the samples that belong to unknown classes.
We take the last layer of sDNN before dense layers as the input features for EVM. Then the open-world classifier notifies classification results to the user if given samples belong to known classes, and gives feedback to the sensor dynamic scheduler. The samples rejected by the classifier (unknown samples) are placed into a queue for manual labeling under supervised learning and for clustering under unsupervised learning.

\subsection{Incremental Learning Process} 
\label{sec:incremental_learning}
There are three criteria that an incremental learning framework must meet for practical applications:
\begin{enumerate}
 \item The model can be trained from a stream of data introducing new classes to the classifier over time.
 \item The inference performance must not be significantly lowered, especially due to the catastrophic forgetting phenomena~\cite{MCCLOSKEY1989109}.
 \item Updating the model must meet the resource requirements of the system since the framework should be able to run on embedded devices
\end{enumerate}

For supervised learning, the samples in the queue are manually labeled and then we can update the classifier.
For unsupervised learning, we need to identify and assign new classes to the unknown samples in the queue via a clustering algorithm before updating the classifier. Since this is a clustering problem where the number of groups is unknown, our framework uses the Finch algorithm~\cite{sarfraz2019efficient} which is a parameter-free clustering method. The Finch algorithm provides different partitioning results for the given samples. Then our framework chooses the partition with the minimum number of clusters and only select clusters with at least a certain number of samples as new classes and label them accordingly~\cite{Jafarzadeh2020OpenWorldLW}. The open-world model updater will incorporate these new classes into our classifier based on EVM.


One advantage of using the EVM-based classifier is that it only needs the extracted features of the new samples to update the classifier without the need to retrain the entire DNN, enabling more effective and efficient incremental learning. Another advantage is controlling the model size by selecting the minimum number of points needed to cover a known class, which is known as the set cover problem~\cite{slav1996coverproblem} and important for incremental learning on resource-constrained devices. 

\subsection{Sensor Scheduler}
\label{sec:dynamic_scheduler}
\blue{While many smart sensing frameworks use a fixed interval, e.g., one second, to pull the sensor data and infer it~\cite{laput2017synthetic,DeepSense}, we aim to develop a dynamic sensor scheduler that determines a flexible sensing and sensor data transmission period between the sensor board and the edge device. 
Our objective is two folds: (i) increase idle time (i.e., the time between two consecutive sensor data transmissions) to reduce the energy consumption of the sensor board and free resources on the edge device for open-world learning tasks, and (ii) ensure the timeliness of inference to avoid unnecessary delays and missing important events. 
To set criteria for the sensing period and satisfactory inference performance, we define a classification latency constraint, $CL_s$, which is the maximum delay that can be tolerated in the detection and classification of an event of a class type $s$ in the environment. In other words, if the edge device is idling longer than $CL_s$ from the moment an event happens, it may miss the event or give unsatisfactory responsiveness to the user.
Therefore, the goal of the sensor scheduler is to determine a sensing period for each class $s$ such that it does not exceed the respective $CL_s$ constraint of that class.
$CL_s$ imposes the following condition to be satisfied:
\begin{equation}\label{eq:CL}
T_{sp} \:-\:T_{e}\:\leq \:CL_s 
\end{equation}
where $T_{sp}$ is the sensing period 
assigned by the scheduler for the current event $e$ in class $s$ (e.g., $T_{sp}=1$ under the conventional fixed-interval sensing frameworks), and $T_e$ is the time interval that the event $e$ from class $s$ lasts until another event of different class $s'$ appears in the environment. Without loss of generality, we assume discrete time and $T_e > 0$.} 

\blue{An ideal scheduler will give $T_{sp}$ equal to $T_e$, leading to zero classification delay. However, in practice, such a scheduler is nearly impossible to implement because each event can run for a different amount of time, making this a challenging problem. 
Therefore, we propose two dynamic scheduling algorithms that determine a suitable $T_{sp}$ in different ways. 
Our first scheduler, class-level period assignment, gives a fixed $T_{sp}$ for each class type, i.e., different classes have different $T_{sp}$ values.  
Our second scheduler is based on reinforcement learning, specifically Q-learning,  
to change $T_{sp}$ values even for the same class of events depending on the current class and the previous actions made.


\begin{algorithm}[tbp]
\small
\DontPrintSemicolon
  \SetKwInOut{Input}{Input}
  \SetKwInOut{Output}{Output}
  \Input{$T_{e}$: All time intervals for class $s$\newline $CL_s$: Classification latency constraint for class $s$}
  \Output{$T_{sp}$: Sensing period for class $s$}
  
  $T_{e} \gets {AscendingSort{(T_{e})}}$\;
  $T_{sp} \gets {FindMinimum{(T_{e})}}$\;
  $L1 \gets T_{sp}$\;
  $L2 \gets {length{(T_{sp})}}$\;
  $i \gets 0$; $j \gets 0$\;
  
  \For{$i \leq L1$}
  {
    \For{$j \leq L2$}
    {
      $n \gets {ceil{(T_{e}[j]/T_{sp})}}$\;
      
      \If{$n \times T_{sp} \geq T_{e}[j]+CL_s$}
      {
        $T_{sp} \gets T_{sp} - 1$\;
        break\;
      }
      $j++$\;
    }
    $i++$\;
  }
  \eIf{$T_{sp} > 1$}
  {
    return $T_{sp}$\;
  }{
    return fail\;
  }
  \caption{Class-level Period Assignment}\label{alg:class_level_sched}
\end{algorithm}

\noindent \textbf{Class-Level Period Assignment (CLPA):} The main idea of this scheduler is to approximate the largest common factor of all periods for a given class $s$ such that each assigned $T_{sp}$ meets the $CL_s$ condition as well as that the sensing period to be greater than one. The reason behind ``approximating'' instead of directly computing the greater common factor is that, when doing the latter, the resulting $T_{sp}$ value will be most likely equal to one for most cases due to the dynamic nature of event occurrences, resulting in similar behavior as the conventional fixed time interval approach.

To mitigate the risk of missing an event, this algorithm splits the ideal sensing interval into $n$ sub-intervals and finds $T_{sp}$ for each sub-interval. This approach modifies Eq.~\eqref{eq:CL} as follows:} 
\begin{equation}\label{eq:CL2}
n\: \times \:T_{sp} \:-\:T_{e}\:\leq \:CL_s
\end{equation}

Alg.~\ref{alg:class_level_sched} gives the pseudo code for the class-level period assignment algorithm. It selects the minimum time interval among all time intervals for a class $s$ as a base sensing period, $T_{sp}$.
The base period is compared with all other time intervals such that the difference to $T_e$ after $n$ cycles does not exceed $CL_s$. 
If the base period $T_{sp}$ does not meet this condition, the algorithm decrements $T_{sp}$ by one and continues the search. 
If $T_{sp}$ becomes one or smaller, it means there is no feasible $T_{sp}$ that satisfies the user's latency requirement and thus the algorithm returns fail.
Then, the user can increase $CL_s$ to increase the search space and rerun the algorithm. In the worst case where no feasible $T_{sp}$ is found for a given class $s$, the user may decide to set $CL_s$ for this class to the minimum value of one, which ensures $T_{sp}$ to be at least two, i.e., $CL_s=1$ and $T_{sp}=2$.\footnote{\blue{With $T_{sp}=2$, the difference between $n \times T_{sp}$ and any positive integer $T_e > 0$ is either 0 or 1, thereby satisfying Eq.~\eqref{eq:CL2}.}} 
Assuming that sensing and classification operations can be completed in one time unit, 
$T_{sp}=2$ still ensures the sensor board and the edge device to be idle for at least half of the sensing period, reducing energy consumption by up to half.\footnote{The actual amount of energy savings in such a worst-case condition $T_{sp}=2$ may vary depending on the idle energy consumption of the board.} The user has the freedom to choose a single $CL_s$ for all events or specify different values for individual events based on the importance and tolerated latency.

\blue{Although this approach is effective in finding $T_{sp}$ values that are larger than one second and can meet the $CL_s$ condition, these periods are fixed for respective class types, which may cause energy waste and excessive classification latency in a changing environment.}

\begin{algorithm}[tbp]
\blue{
\small
\DontPrintSemicolon
\SetKwInOut{Input}{Input}
\SetKwInOut{Output}{Output}

\Input{$state$: Current state represented by a tuple $(class, prev\_T_{sp})$, where $class$ is the class type of the current event's class type and $prev\_T_{sp}$ is the previous action that made $class$. \newline
$action$: New action to take, i.e., a new $T_{sp}$ value. 
}
\Output{$next\_state$: Next state based on the $action$ taken. \newline
$reward$: Positive or negative reward value based on the $action$ taken.}

$ class, prev\_T_{sp} \gets state $ \;
$ T_{sp} \gets action $\;
Update $T_{ideal}$ based on time elapses\;  
\eIf(/* New event arrived */){$ T_{sp} \geq T_{ideal} $}{
    $ class \gets $ Obtain new event's class\;
    \eIf{ $Cr_1: T_{sp} - T_{ideal} \leq CL_s $}{
        return $reward = weight_{p1}$\;
    }{
    return $reward = weight_{n1}$\;
    }
}
(/* Current event continues */){
    $T_{ideal} \gets T_{ideal} - T_{sp} $\; 

    \eIf{$Cr_2: T_{sp} \geq prev\_T_{sp} $}{
            return $reward = weight_{p2}$\;
        }{
         return $reward = weight_{n2}$\;
        }

}
$ next\_state \gets (class, T_{sp})$ \;
return $(next\_state, reward)$
\caption{QL $take\_action()$ function}\label{alg:take_action}
}
\end{algorithm}

\begin{algorithm}[tbp]
\small
\DontPrintSemicolon
\SetKwInOut{Input}{Input}
\SetKwInOut{Output}{Output}
\blue{
\Input{$mode$: $update$ if updating the scheduler at runtime on the edge device, $null$ otherwise\newline
$old\_Q$: Existing Q-table to update ($\emptyset$ if $mode= null$)\newline
$\epsilon$: Parameter to determine whether to explore the action space or exploit learned weights \newline
$\theta$: Threshold to stop the training loop \newline
$n_{success}$: \# of consecutive episodes that satisfy $\theta$ \newline
$(\alpha,\gamma)$: Training hyper-parameters\newline
$n_{episodes}$: \# of training episodes }
\Output{$Q_{table}$: A 2-D matrix for actions' weight for each state in the model ($\#states \times \#actions$)}
$Q_{table} \gets old\_Q$ /* Initialize the Q-table */\;
$i\gets 0; cur\_avg\_penalty \gets 0; penalty\_count\gets 0$; \;
\For{$i \leq n_{episodes}$}{\label{line:train_QL_iter}
    $ done \gets   false$; $state \gets 0$; \;
    $ prev\_avg\_penalty \gets cur\_avg\_penalty$\;
    \While{$done = false$}{
        \eIf{$random(0,1) \leq \epsilon$}{
            $action \gets random(1,100)$\;\label{line:train_QL_action_random}
        }{
            $action \gets max(Q_{table}[state])$\label{line:train_QL_action_table}
        }
        $(next\_state, reward) \gets take\_action(action,state)$ /* Alg.~\ref{alg:take_action} */\;
        Update $Q_{table}[next\_state, reward]$ using
        $Q_{n} = (1-\alpha)Q_{n-1}+\alpha(reward+\gamma Q_{n+1}) $\;\label{line:train_QL_update_Q_table}
        $state \gets next\_state$
    
        \If{$reward\textless0 $}{ 
            $penalty\_count++$\;
        }        
        \If{all data points are covered}{       
          $ done \gets true$ \;
          $cur\_avg\_penalty \gets penalty\_count / i $\;
        }
    }
    \If{$mode$ = $update$}{      
            
        \eIf{$abs(cur\_avg\_penalty-prev\_avg\_penalty) \leq \theta$}{    
            $success\_episodes++$\;
        }{
            $success\_episodes \gets 0 $\;
        }       
        \If{$success\_episodes \geq n_{success}$}{\label{line:train_QL_success}
            \textbf{return} $Q_{table}$ /* Terminate update */ \;
        }        
    }

}
\textbf{return} $Q_{table}$ /* Finish training */\;

\caption{QL-model training and updating loop}
\label{alg:train_QL}
}
\end{algorithm}

\blue{
\noindent \textbf{Q-Learning-Based Scheduler (QLBS):}
We propose a reinforcement learning-based scheduler, specifically using Q-learning (QL). The purpose of the QL model is to determine a suitable sensing period, $T_{sp}$, before entering an idle phase, given the state of the current event and the previous $T_{sp}$ values. To train this model, we consider two criteria, $Cr_{1}$ and $Cr_{2}$. First, $Cr_{1}$ is to compare $T_{sp}$ to $T_{ideal}$, which is the time it takes for the current event of class $s$ to change to the next event of class $s'$ ($s \ne s'$). 
Notice that whenever determining a new $T_{sp}$ value, $T_{ideal}$ also gets updated since the wall clock time has elapsed and less time is left until the current event changes from $s$ to $s'$.
The first criterion $Cr_{1}$ is thus represented by the following condition:   
\begin{equation}\label{eq:CL3}
T_{sp} \:-\:T_{ideal}\:\leq \:CL_s
\end{equation}
Here, we want the scheduler to pick $T_{sp}$ from a specific range of values, e.g., $T_{sp}$ $\in$ [0$-$100] seconds, to limit our state space. This range should be set close to or greater than possible $T_{ideal}$. The second criterion $Cr_{2}$ is to choose $T_{sp}$ that is larger than or equal to the previous $T_{sp}$ ($prev\_T_{sp}$), i.e., $T_{sp}\ge prev\_T_{sp}$, so as to increase the idle time until the class of the current event changes. Each criterion is parameterized by the ratio between a reward ($weight_{p}$) and a penalty ($weight_{n}$), applied depending on whether its respective condition is met. Alg.~\ref{alg:take_action} shows detailed steps for deciding a penalty or reward when an action is taken by our QL model training, with the consideration of the two aforementioned criteria. The ratio values for the two criteria and their relative degree affect the performance of the proposed scheduler and can be configured based on the user's preference, e.g., increasing energy efficiency or lowering classification delay. 
}


\blue{Alg.~\ref{alg:train_QL} gives our algorithm for both training the QL model offline and updating it at runtime. In each training episode
, the algorithm goes through a sequence of actions, where each action is a $T_{sp}$ value to take in the current state, and updates the Q-table until it 
finishes a given number of episodes for offline training, or the difference in average penalty between two episodes is less than or equal to a threshold $\theta$ when updating the QL model at runtime.  
The action ($T_{sp}$) is picked either: (i) randomly to explore a possible value that can improve the model (line \ref{line:train_QL_action_random}), or (ii) by selecting the one with the highest weight in the Q-table to take the best possible action (line \ref{line:train_QL_action_table}). The selection between (i) and (ii) is controllable by the input parameter $\epsilon$. Once the action is picked, the next state and the corresponding reward are obtained using the $take\_action()$ function (Alg.~\ref{alg:take_action}), and the Q-table is updated accordingly (line \ref{line:train_QL_update_Q_table}). This procedure continues until it covers all timestamps of the given training dataset.}

\blue{The Q-table is initially fully trained on an external server since it requires a large dataset and a high number of episodes to produce a good model. It is obviously not a viable option to fully re-train the entire QL model from scratch on an edge device. Hence, our algorithm allows the edge device to update an existing model by performing the training sequence only until the difference in the average penalty between two training episodes
 is less than or equal to a given threshold $\theta$ for a certain number of consecutive episodes $n_{success}$. 
With this, the edge device can run Alg.~\ref{alg:train_QL} when it is idling and has collected a sufficient amount of new event intervals ($T_e$).
More details on how different threshold values affect the scheduler performance are presented in Sec.~\ref{sec:scheuler_results}.}

\begin{algorithm}[tbp]
\blue{
\small
\DontPrintSemicolon
\SetKwInOut{Input}{Input}

\Input{
$update\_mode: $ $classifier$ to update the inference model or $scheduler$ to update the scheduler.\newline
$N_{u}$: \# of samples of the newly discovered class. \newline
$S_{u}$: Samples from the newly discovered class. \newline
$T_{sp}$: Current sensing period (i.e., available idle time). \newline 
$scheduler\_type:$ CLPA or QLBS}

\uIf{$ update\_mode = classifier$}{    

    $N_{old} \gets 0$\;
    \While{$N_{u} \le 0$}{
        \If{$T_{sp} \geq T_{min}$}{
            $N_{ST} \gets {ComputeSamplesToTrain{(T_{sp})}}$\;
            ${UpdateModel{(S_{u}[N_{old}:N_{ST}])}}$\;
            \eIf{$N_{u} \geq N_{ST}$}{
                $N_{old} \gets N_{ST}$\;
                $N_{u} \gets N_{u} - N_{ST}$\;
            }{
                \textbf{return} success\;
            }
        }    
        \textbf{return} fail  /* wait for next $T_{sp}$ */ \;
    }
}
\ElseIf{$update\_mode = scheduler$}{
    Get new $ T_e$ intervals for each class\;
    /* Update Class-level Period Assignment (CLPA) */\;
    \uIf{$ scheduler\_type = \mathrm{CLPA}$} {
        \textbf{call} Alg.~\ref{alg:class_level_sched}\;
    }
    /* Update QL-Based Scheduler (QLBS) */ \;
    \ElseIf{$ scheduler\_id = \mathrm{QLBS}$}{ 
        $ training\_mode = on\_edge$\;
        \textbf{call} Alg.~\ref{alg:train_QL}\;
    }
}

\caption{Classifier/Scheduler Updater
}
\label{alg:model_update_scheduler}
}
\end{algorithm}

\blue{

\subsection{Classifier and Scheduler Updater
}
\label{sec:classifier_updater}
Alg.~\ref{alg:model_update_scheduler} outlines our proposed procedure to update the classifier of the inference model and the sensor schedulers in a resource-constrained edge device.
The updater can be set to update either the inference model or the sensor scheduler one at a time by the input parameter $update\_mode$. 

In the case of updating the inference model ($update\_mode=classifier$), it can be updated once the number of unknown samples clustered into a new class exceeds the minimum requirement. However, due to the fact that our proposed framework is designed for resource-constraint edge devices, Alg.~\ref{alg:model_update_scheduler} only updates the model partially with a certain number of samples from the newly discovered class based on the sensing period $T_{sp}$ currently set by the sensor scheduler. 
We fit a polynomial ($ComputeSamplesToTrain$) to compute the number of samples that can be trained during the idle time available in the current period, given that it meets the minimum average time to train at least one sample ($T_{min}$).}
Note that $T_{sp}$ may not satisfy the condition ({$T_{sp} \geq T_{min}$}). Therefore, different approaches to choose $T_{sp}$ can be used to extend the length of the sensor idling time without significantly compromising the latency performance of the scheduler, as can be seen in Sec.~\ref{sec:scheuler_results}.
After the model is updated, the update deploys the new open-world classifier to enable the framework to recognize the newly learned classes and discover new classes from the upcoming time-series data from sensor boards. 

\blue{If $update\_mode$ is set to $scheduler$, the updater first checks if there are sufficient event intervals ($T_{e}$) for each class, e.g., five $T_{e}$ samples per class. The input parameter $scheduler\_type$ determines which sensor scheduler algorithm to be updated. If it is ``CLPA'', Alg.~\ref{alg:class_level_sched} will be called to recompute a new $T_{sp}$ for each class. If $scheduler\_type$ is set to ``QLBS'', $training\_mode$ is set to $update$ to ensure that Alg.~\ref{alg:train_QL} runs only until the average penalty converges to the threshold $\theta$ since it is infeasible to fully re-train the QL model on the edge device as we discussed earlier. Once done, the updated $T_{sp}$ for each class given by Alg.~\ref{alg:class_level_sched} or the updated Q-table will be used for sensor scheduling.}

\section{Evaluation}
There are two essential performance metrics to evaluate the overall performance of our proposed framework on resource-constrained edge devices: (i) classification and incremental learning, and (ii) latency and energy consumption. For classification and incremental learning, we first compare the inference and efficiency performances of our proposed sDNN to DeepSense. After that we investigate the effectiveness of OpenSense in both supervised and unsupervised settings and compare it to other methods. We analyze the efficiency of each algorithm as an open-world learner by comparing the execution time of different tasks and evaluate our design choices on an embedded device. Lastly, we show the latency and energy consumption performances for the proposed sensor scheduler approaches. In this section, we also provide more details on the evaluation platforms and the dataset.

\subsection{Evaluation Platforms}\label{sec:eval_platform}

\green{OpenSense was implemented on Raspberry Pi 4 Model B to evaluate the proposed framework's runtime performance and overhead on a realistic embedded edge device. 
For comparative experiments with other existing approaches, we also used a Linux-based machine equipped with Intel i7-8650U, 16GB memory, and a dedicated NVIDIA GeForce GTX 1060 GPU with 6GB VRAM. The reason behind the use of the Linux machine is that the existing approaches were not originally developed for use on edge devices and they need high computational power from GPU. With the experiment results from both platforms, the reader can understand the performance trends of our framework across various hardware settings.}

\subsection{Datasets}

We used two datasets, HHAR and PAMAP2, that have been widely used in the literature of time-series human activity recognition. These datasets consist of time-series human activities data captured by different sensors and classified into distinct classes based on the activities of the wearer. 
The details of each dataset and preprocessing are as follow.

\noindent \textbf{HHAR:} The heterogeneous human activity recognition (HHAR)~\cite{Stisen2015sdad} dataset records 6 activities 
performed by 9 different users. The data is collected using the accelerometer and gyroscope embedded on smartphones and smartwatches. We followed the preprocessing suggested by DeepSense~\cite{DeepSense} to reproduce similar results for evaluation: time-series data is divided into 5-seconds non-overlapping windows, each window is divided into 0.25 second segments, and FFT is applied to each segment to compute the frequency response. The number of samples after preprocessing is around 120K and we divided them into training (70\%), validation (10\%) and testing (20\%) sets. 

\noindent \textbf{PAMAP2:} The physical activity monitoring data set (PAMAP2)~\cite{6246152} is collected using 3 inertial measurement units (IMU) sampled at 100Hz, each capturing temperature, acceleration, gyroscope and manometer, attached at wrist, chest and ankle of the user, and a heart sensor at 9Hz (total of 13 sensors data). The data is recorded by monitoring 18 activities
performed by 9 users. The time-series data is divided into 1-second windows, each window is divided into ten segments each is 0.1 second. FTT is applied to each segment to calculate the frequency response. The number of samples after preprocessing is around 27K divided into training (60\%), validation (20\%) and testing (20\%) sets since the number of samples is much smaller per class compared to HHAR.

Since both datasets are imbalanced, we perform \green{a hybrid undersampling and oversampling technique} for each class in each dataset set to ensure that each class weighs equally to the performance, especially in the incremental learning experiments.

\blue{
\noindent \textbf{Kitchen Dataset:}
Since the above two datasets are relatively sparse in terms of event occurrences, 
we used a simulated dataset based on the actual data of kitchen events collected using the sensor board in Sec. \ref{sec:sensor_board} to evaluate the performance of our proposed sensor scheduling algorithms over a long-term continuous deployment environment. 
The dataset contains six different classes of events that can be captured by commodity IoT sensors from a user environment. The six classes include \green{[~`0: None', `1: Microwave', `2: Kettle', `3: Faucet', `4: Waste Disposer', `5: Vent Fan'~]. }The training set contains 7,000 seconds of sensor reading each second, randomly generated with different interval for each event. The testing dataset is randomly generated at each test, with a range of varying length between 5,000-8,000 seconds.}

\subsection{Learning Algorithms to Compare}
We evaluate OpenSense and compare it to 2 learning algorithms in both supervised and unsupervised settings.
\newline
\noindent \textbf{na\"{\i}ve approach (NA):} The based DNN model is updated using only the data coming in the data stream without accessing previously trained samples.
\newline
\noindent \textbf{Fixed-representation Class Incremental Learning (FRCI)~\cite{Rebuffi2017iCaRLIC}:} In this algorithm the DNN model with the new samples in the data-stream but with the addition of exemplars learned based on the Nearest Mean Class to avoid Catastrophic Forgetting while incrementally learn without imposing an overhead on the system.
\subsection{Results: Classification and Learning Performance}
\noindent \textbf{DNN Base Models and EVM:} The base DNN is very important to the proposed framework for building a sustainable features extractor, EVM model and a classifier for other incremental learning approaches. Therefore, we compare sDNN with a state-of-the-art architecture, DeepSense. 
The experiment is conducted on 2 different datasets. For DeepSense, we implemented it in Keras and successfully reproduced the results in the original work. We then compare it with sDNN, implemented on Keras as well, to produce consistent evaluation. We used fixed parameters in our comparison and used accuracy and F1-macro scores for evaluation, similar to DeepSense~\cite{DeepSense}. We also built an EVM model based on sDNN to evaluate it as a classifier. We used fixed parameters as well for EVM training in all experiments. We performed cross-validation in the early stage of development and found out that the following EVM parameters: tail-size = 100, cover threshold = 0.7 and distance multiplier = 0.4 produce acceptable results across the two time-series dataset we used. In this set of experiments, all models are trained with all classes as known classes and the training is performed on batches with all samples in the training dataset. \green{As mentioned in Sec.~\ref{sec:eval_platform}, we perform comparative experiments with existing methods on the Linux machine, and then present the performance of OpenSense on Raspberry Pi at the end of this subsection.}

\begin{figure}[t!]
\centering
\begin{subfigure}{.5\textwidth}
 \centering
 \includegraphics[width=\linewidth]{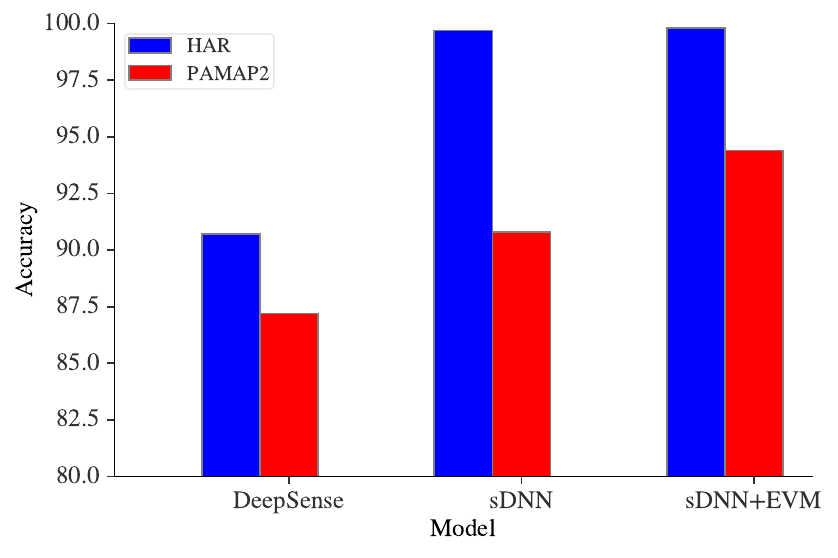} 
\end{subfigure}
\caption{Inference performance of DeepSense, sDNN, sDNN+EVM}
\label{fig:acc_sdnn}
\label{fig3}
\end{figure}

\begin{table}[t!]
\centering
\footnotesize
\caption{\green{Precision, recall, and F1-score of DeepSense, sDNN, sDNN+EVM}}
\label{tab:perf_dnn}
\resizebox{\columnwidth}{!}{\begin{tabular}{|l|ccc|ccc|}
\hline 
Dataset   & \multicolumn{3}{c|}{HAAR}                                                & \multicolumn{3}{c|}{PAMAP2}                                             \\ \hline
Metric    & \multicolumn{1}{c|}{Recall} & \multicolumn{1}{c|}{Precision} & F1-Score & \multicolumn{1}{c|}{Recall} & \multicolumn{1}{c|}{Precision} & F1-Score \\ \hline
DeepSense & \multicolumn{1}{c|}{90.6\%} & \multicolumn{1}{c|}{91.1\%}     & 90.7\%   & \multicolumn{1}{c|}{86.1\%} & \multicolumn{1}{c|}{86.7\%}    & 85.9\%   \\ \hline
sDNN      & \multicolumn{1}{c|}{99.7\%} & \multicolumn{1}{c|}{99.7\%}     & 99.7\%   & \multicolumn{1}{c|}{90.8\%} & \multicolumn{1}{c|}{91.2\%}    & 90.6\%   \\ \hline
sDNN+EVM  & \multicolumn{1}{c|}{99.8\%} & \multicolumn{1}{c|}{99.8\%}     & 99.8\%   & \multicolumn{1}{c|}{94.5\%} & \multicolumn{1}{c|}{94.3\%}    & 94.4\%   \\ \hline
\end{tabular}}
\end{table}

\begin{table}[t!]
\centering
\footnotesize
\caption{DNN Model Efficiency}
\label{tab:eff_sdnn}
\begin{tabular*}{\linewidth}{@{\extracolsep{\fill}}|l|cc|cc|}
\hline
Dataset & \multicolumn{2}{c|}{HHAR} & \multicolumn{2}{c|}{PAMAP2} \\ \hline
DNN Model & \multicolumn{1}{c|}{DeepSense} & sDNN & \multicolumn{1}{c|}{DeepSense} & sDNN \\ \hline
\begin{tabular}[c]{@{}l@{}}\#epochs to\\ converge\end{tabular} & \multicolumn{1}{c|}{100} & 10 & \multicolumn{1}{c|}{150} & 50 \\ \hline
\begin{tabular}[c]{@{}l@{}}average execution \\ time \textbackslash epoch
\end{tabular} & \multicolumn{1}{c|}{37 sec} & 16 sec & \multicolumn{1}{c|}{18 sec} & 3 sec \\ \hline
speedup \textbackslash epoch & \multicolumn{2}{c|}{x2.3} & \multicolumn{2}{c|}{x6} \\ \hline
\begin{tabular}[c]{@{}l@{}}\green{total training}\\ \green{time (Linux machine)}\end{tabular} & \multicolumn{1}{c|}{\begin{tabular}[c]{@{}c@{}}61 min \\ 40 sec\end{tabular}} & \begin{tabular}[c]{@{}c@{}}2 min \\ 31 sec\end{tabular} & \multicolumn{1}{c|}{45 min} & \begin{tabular}[c]{@{}c@{}}2 min \\ 30 sec\end{tabular} \\ \hline
\end{tabular*}
\end{table}
We can see that the proposed DNN outperforms DeepSense on both HHAR and PAMAP2 datasets in Fig.~\ref{fig3} and Table~\ref{tab:perf_dnn}. We can also see that sDNN can produce robust features, when freezing the last layer, for the EVM model which slightly provide better accuracy and F1-score than the sDNN model on the same testing sets. In addition to that, sDNN is faster than DeepSense by up to x2.3 and x6 per epoch for HHAR and PAMAP2, respectively as shown in Table ~\ref{tab:eff_sdnn}. We can see that sDNN converges much faster than DeepSense on both datasets. Faster DNN training can allow for more efficient representation update in case we consider updating the base DNN in our framework, and in other supervised incremental learning approaches as well. When implementing DeepSense, each sensor in the sensor board will add three individual convolutional layers to the framework leading to a total of 12 and 45 layers for HHAR and PAMAP2, respectively compared to a fixed 5 layers in sDNN, two convolutional layers, two LSTM layers and an output layer. We assumed that due to the increased complexity of DeepSense, the training data is over-fitting the model, especially with PAMAP2 dataset that has limited number of samples and requires longer training time to converge. 

\noindent \textbf{Supervised Incremental Learning:} In the supervised incremental learning setup, the data-stream is divided into equally batched samples with a specific number of known classes. The known classes are incoming as an increment of 2 classes per data-stream for HHAR and an increment of 3 classes for PAMAP2. For OpenSense, the base DNN is trained with the first batch of data, 2 and 3 classes from HHAR and PAMAP2 respectively. The purpose of this experiment is to compare the performance of OpenSense to NA and FRCI algorithms, in case we manually label new classes and incrementally update the model. We used accuracy as a metric to compare each algorithm since all samples are labeled. 

\begin{figure}[t!]
\centering
\begin{subfigure}{.5\textwidth}
 \centering
 \includegraphics[width=\linewidth]{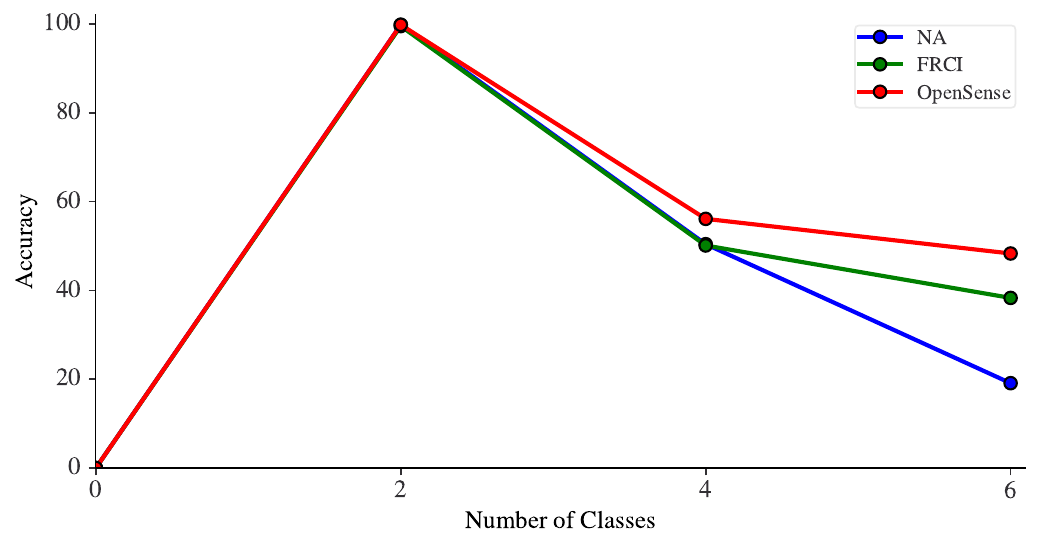} 
 \caption{HHAR}
 \label{fig:sub-first}
\end{subfigure}
\begin{subfigure}{.5\textwidth}
 \centering
 \includegraphics[width=\linewidth]{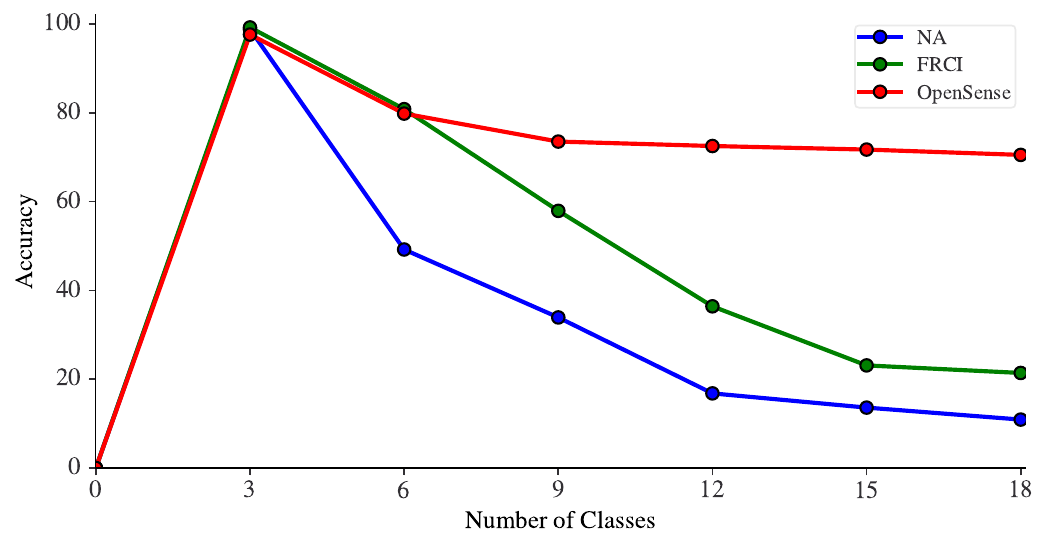} 
 \caption{PAMAP2}
 \label{fig:sub-second}
\end{subfigure}
\caption{Comparing the accuracy of the na\"{\i}ve approach (NA), FRCI and OpenSense for each step in a supervised incremental learning set-up}
\label{fig4}
\end{figure}

We can see in Fig.~\ref{fig4} that all algorithms achieve around 99\% and 95\% accuracy for HHAR and PAMAP2, respectively, with both the na\"{\i}ve approach, noted as NA, and FRCI slightly better accuracy than OpenSense in the initial training phase. As more data coming in the stream, we can clearly see that OpenSense outperforms NA and FRCI on both datasets since OpenSense uses the features extracted by the base DNN model, in which the features representation is fixed for all class. The EVM model of OpenSense then use these features with labels and incrementally assign new EVM vectors to represent each class.

For the na\"{\i}ve approach and FRCI, their accuracy significantly drops to 50\% in HHAR dataset with much better performance for FRCI in PAMAP2 dataset closely to OpenSense. NA accuracy keeps falling as it incrementally learning to 34.8\% and 10.9\% for HHAR and PAMAP2, respectively. We see that during testing after each increment in NA that it almost completely forgets any classes that are learned in the previous increment due the catastrophic forgetting problem. Although FRCI performance is similar to OpenSense on HHAR dataset, we see that, similar to NA, it reaches to a low 20.4\% accuracy by the end of the incremental learning of PAMAP2 dataset. \green{Table~\ref{tab:perf_super} shows recall, precision, and F1-score for the last learning increment. It shows that OpenSense outperforms NA and FRCI.}  Updating the DNN model with a no or limited number of samples from previously learned classes leads to a degrade in its inference performance, making OpenSense more suitable for incremental learning.

\begin{table}[t!]
\centering
\footnotesize
\caption{\green{Precision, recall, and F1-score of supervised incremental learning}}
\label{tab:perf_super}
\resizebox{\columnwidth}{!}{\begin{tabular}{|l|ccc|ccc|}
\hline
Dataset   & \multicolumn{3}{c|}{HAAR}                                                & \multicolumn{3}{c|}{PAMAP2}                                             \\ \hline
Metric    & \multicolumn{1}{c|}{Recall} & \multicolumn{1}{c|}{Precistion} & F1-Score & \multicolumn{1}{c|}{Recall} & \multicolumn{1}{c|}{Precision} & F1-Score \\ \hline
NA        & \multicolumn{1}{c|}{32.9\%} & \multicolumn{1}{c|}{9.6\%}      & 14.1\%   & \multicolumn{1}{c|}{10.8\%} & \multicolumn{1}{c|}{2.1\%}     & 3.1\%    \\ \hline
FRCI      & \multicolumn{1}{c|}{40.8\%} & \multicolumn{1}{c|}{25.1\%}     & 28.9\%   & \multicolumn{1}{c|}{21.2\%} & \multicolumn{1}{c|}{12.2\%}    & 10.7\%   \\ \hline
OpenSense & \multicolumn{1}{c|}{50.8\%} & \multicolumn{1}{c|}{47.9\%}     & 43.4\%   & \multicolumn{1}{c|}{70.6\%} & \multicolumn{1}{c|}{77.8\%}    & 71.9\%   \\ \hline
\end{tabular}}
\end{table}

\noindent \textbf{Open-World (Unsupervised) Incremental Learning:} In the unsupervised incremental learning setup, only the first batch of the training dataset consists of known classes. 
Unlike the supervised setup, we assume that initially we have a large number of known classes and incrementally discover new classes from the data-stream.
Since HHAR dataset has a limited number of classes, namely six, it is not sufficient for the evaluation of this experiment.
Therefore, we only performed this experiment on PAMAP2 where the initial DNN model is trained on the dataset of 9 known classes and the remaining class are incoming as 3 unknown classes in each data-stream. Since OpenSense is designed for open-world problem, we added a novel class detector and unsupervised clustering algorithm for the na\"{\i}ve approach and FRCI by setting a threshold on the output of the DNN, such that any sample with a probability less than the threshold will be rejected as an unknown sample. The threshold is chosen based on the probability history of previously discovered classes by taking the mean of the maximum probability of each prior sample. The rejected samples are collected in a queue to cluster them into novel classes and assign a label for each new class. 

\green{Unlike the supervised incremental learning case, conventional performance metrics, such as accuracy, precision, recall, and F1-score, cannot be used in the unsupervised incremental learning environment because we do not have labels for newly-arriving unknown classes~\cite{Jafarzadeh2020OpenWorldLW}. Hence, we use the Open-World Metric (OWM) proposed in \cite{Jafarzadeh2020OpenWorldLW}, which is defined as follows:}
\[OWM = \frac{N_{KK}\cdot Acc(X_{KK})\:+\:N_{UU}\cdot B3(X_{UU})}{N_{KK}+N_{KU}+N_{UK}+N_{UU}}\:\:\:(2)\]
where $X$ is the targeted dataset, $N$ refers to the number of samples in the corresponding category, the subscripts $K$ and $U$ are abbreviations for known and unknown, respectively, e.g., $N_{KU}$ stands for known samples that are misclassified as unknown, $Acc()$ is the accuracy for the known data, and $B3()$ is the B3 score~\cite{baldwin1998description} for the unknown data.

\begin{figure}[t!]
\centerline{
\includegraphics[width=0.5\textwidth]{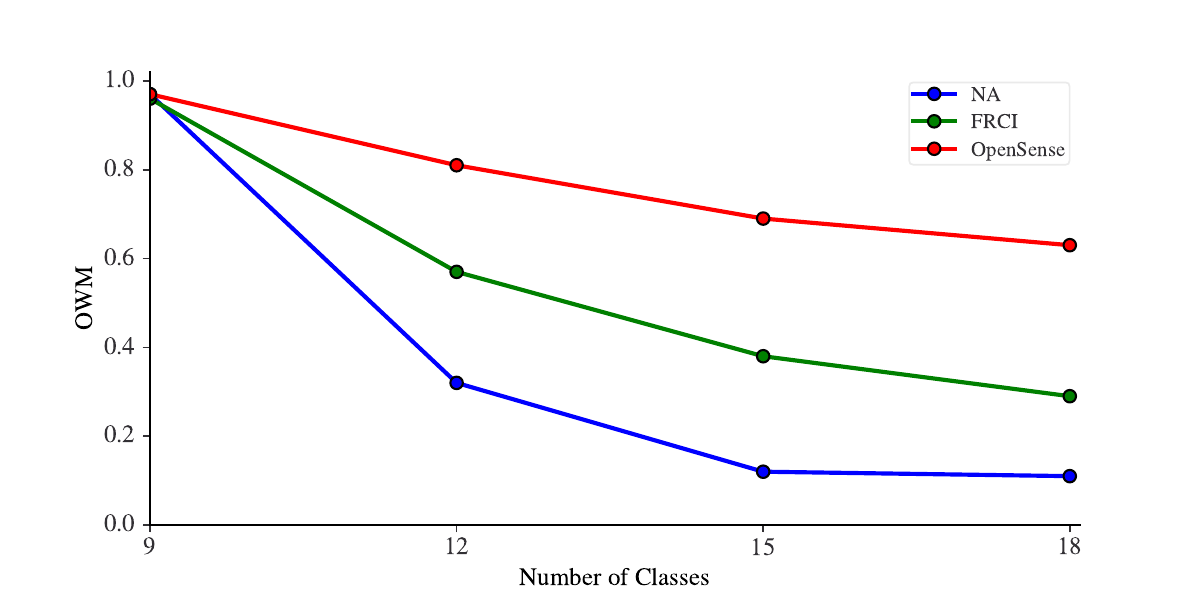}
}
\caption{Comparing the performance of NA, FRCI and OpenSense in an Open-world set}
\label{fig5}
\setlength{\belowcaptionskip}{-50pt}
\end{figure}

Fig~\ref{fig5} shows that all algorithms are very well trained on the initial set with around 95\% accuracy, since all classes are known at this stage. We can see that OpenSense outperforms the other approaches with large margins (0.48 and 0.3) compared to NA and FRCI. The OWM score slowly decreases, with a rate of 0.13 scores, as more unknown samples are discovered by OpenSense. The key to this performance is the ability of EVM-based classifier to reject unknown samples. This allows the clustering algorithm to produce clusters that actually represent unknown classes and update the model accordingly. 

Since the na\"{\i}ve approach suffers from catastrophic forgetting, it is expected that its performance will degrade as more unknown samples are introduced. The NA open-world score was 0.11 after it incrementally exposed to all samples.
Although it performs better than NA, FRCI still struggle to incrementally adapt new classes with an average decrease of 0.23 scores as more unknown classes are coming. Moreover, the number of discovered classes are only 3 and 5 out of 9 classes for NA and FRCI, respectively, wherein OpenSense discovered 9 new classes.

There are many factors, in addition to the fact that most incremental learning works are developed in a supervised approach, that contribute to the expected low performance. One factor is due to the fact that the current unsupervised clustering algorithm, Finch, is unable to cluster the preprocessed time-series data from different classes, while using the features extracted from the base DNN provides better representation to the time-series data. Another important factor is the ability of classifier to correctly reject unknown samples. Even if the features representation is used in the clustering algorithm, it will fail to produce the right number of clusters, or clusters that corresponds to the right label if the rejected samples actually belong to known classes which will lead to confusion to the DNN when updating with new classes.

\begin{table}[t!]
\centering
\footnotesize
  \caption{\green{Inference and incremental learning performance of OpenSense on Raspberry Pi using PAMAP2}}
  \label{tab:perf_pi}
    {\begin{tabular}{|l|c|c|c|c|}
    \hline
    \multicolumn{5}{|c|}{Inference performance of sDNN+EVM} \\ \hline
    Metric                & Accuracy     & F1 & Precision & Recall      \\ \hline
    sDNN+EVM                & 95.8\%            & 94.9\%         & 94.8\%  &95.1\%       \\ \hline 
    \hline
    \multicolumn{5}{|c|}{Inc. learning performance of OpenSense in an Open-world set} \\ \hline
    \#Classes            & 9             & 12        & 15        & 18    \\ \hline
    OWM            & 0.97            & 0.88       & 0.76        & 0.72    \\ \hline           
    \end{tabular}}
\end{table}

\begin{table*}[t!]
\centering
\footnotesize
\caption{The average execution time for different tasks in the
Open-World (Unsupervised) Incremental Learning experiment}
\label{tab:exec_time}
\begin{tabular}{|l|ccc|cc|c|}
\hline
\multirow{2}{*}{Task} & \multicolumn{3}{c|}{Inference} & \multicolumn{2}{c|}{Learning} & \multirow{2}{*}{Total Session Time} \\ \cline{2-6}
 & \multicolumn{1}{c|}{Feature Extraction} & \multicolumn{1}{c|}{Classification} & Queuing & \multicolumn{1}{c|}{Clustering} & Model Updating & \\ \hline
na\"{\i}ve Approach & \multicolumn{1}{c|}{0.5 Sec} & \multicolumn{1}{c|}{12 mSec} & 16 $\mu$Sec & \multicolumn{1}{c|}{0.46 Sec} & 17.4 Sec & 67.8 Sec \\ \hline
FRCI & \multicolumn{1}{c|}{0.47 Sec} & \multicolumn{1}{c|}{35 mSec} & 22 $\mu$Sec & \multicolumn{1}{c|}{0.27 Sec} & 16.5 Sec & 64.5 Sec \\ \hline
OpenSense & \multicolumn{1}{c|}{0.49 Sec} & \multicolumn{1}{c|}{31 mSec} & 18 $\mu$Sec & \multicolumn{1}{c|}{0.13 Sec} & 0.92 Sec & 6.1 Sec \\ \hline
\end{tabular}
\end{table*}

\green{\noindent \textbf{OpenSence Learning Performance on Edge Device: } We now present the inference and incremental learning performance of OpenSense on Raspberry Pi in Table~\ref{tab:perf_pi}. For inference, we evaluated the accuracy of the Open-world classifier (sDNN+EVM) in a supervised setting, where the based model is trained with the labeled dataset (PAMAP2). For incremental learning, we evaluated the entire OpenSense framework in an unsupervised setting, where the initial model is trained with 9 labeled classes. The upper half of Table ~\ref{tab:perf_pi} shows the inference performance of sDNN+EVM on Raspberry Pi, which is similar to the results in Fig.~\ref{fig:acc_sdnn} and Table~\ref{tab:perf_dnn} on the Linux machine. The lower half of the table shows the OWM at each increment that includes samples from 3 unlabeled classes. The framework successfully recognizes these samples and updates the model accordingly on the edge device, thereby achieving inference performance similar to Fig.~\ref{fig5}.}

\subsection{Results: Execution time of Learning Tasks}
\noindent \textbf{Execution Time of Incremental Learning:} We compare the execution time of different tasks in the framework for the 3 incremental algorithms in the open-world learning experiment conducted above, as shown in Table ~\ref{tab:exec_time}. The purpose of this experiment is to demonstrate that our design choices does not only outperform the na\"{\i}ve approach and FRCI in unsupervised incremental learning but also in latency. We compare the execution time of the essential tasks in the framework including preparing the samples for inference, novel class detection, queuing unknown samples, clustering and incrementally updating the model. 
We also compare the overall execution time needed to incrementally discover and adapt new classes for a complete dataset for each algorithm. We measure the execution time of each task in Seconds (s) \green{on the Linux machine. The execution time of OpenSense on Raspberry Pi is presented at the end of this subsection.} 

All experiments are conducted on the same dataset and with 100 samples per batch for the inference task. Note that the number of samples may differ in subsequent tasks depending on classification results. For example, the number of samples in the queue depends on how many samples are classified as unknown by the classifier in the prediction task. 
The last column in the table represents the total time consumed during a complete training session. It takes on average 6 seconds for OpenSense to finish updating the model with all discovered classes from the training dataset, faster by at least x10.5 from NA and FRCI. 
We see that the features extraction latencies are similar across all algorithms since they are based on the same DNN architecture, and lower classification and queuing latencies for the na\"{\i}ve approach compare to FRCI and OpenSense.

In the learning task, NA requires more time for clustering than FRCI and OpenSense due to the fact that the na\"{\i}ve approach forget the classes previously learned leading to a greater rejection rate of unknown samples than other algorithms, wherein OpenSense has lower clustering latency because it has a higher accuracy in predicting unknown samples. As for the updating the model with novel classes, we see that NA and FRCI require in average around 17 and 16 seconds, respectively, 18x slower than OpenSense. Since other algorithms do not need additional tasks before updating the model, we included the task of finding exemplars in the model updating task for FRCI. 

Interestingly, we can update the EVM model regardless of how many samples we have in one epoch making this approach even more suitable for resource-constrained systems. This allow us to process samples periodically as they are coming from the data-stream and choose to update the model when there is enough time given by the dynamic schedulers. We see that in average, it takes less than a second to update the EVM model on a machine with GPU support. 

\begin{figure}[t!]
\centerline{\includegraphics[width=0.5\textwidth]{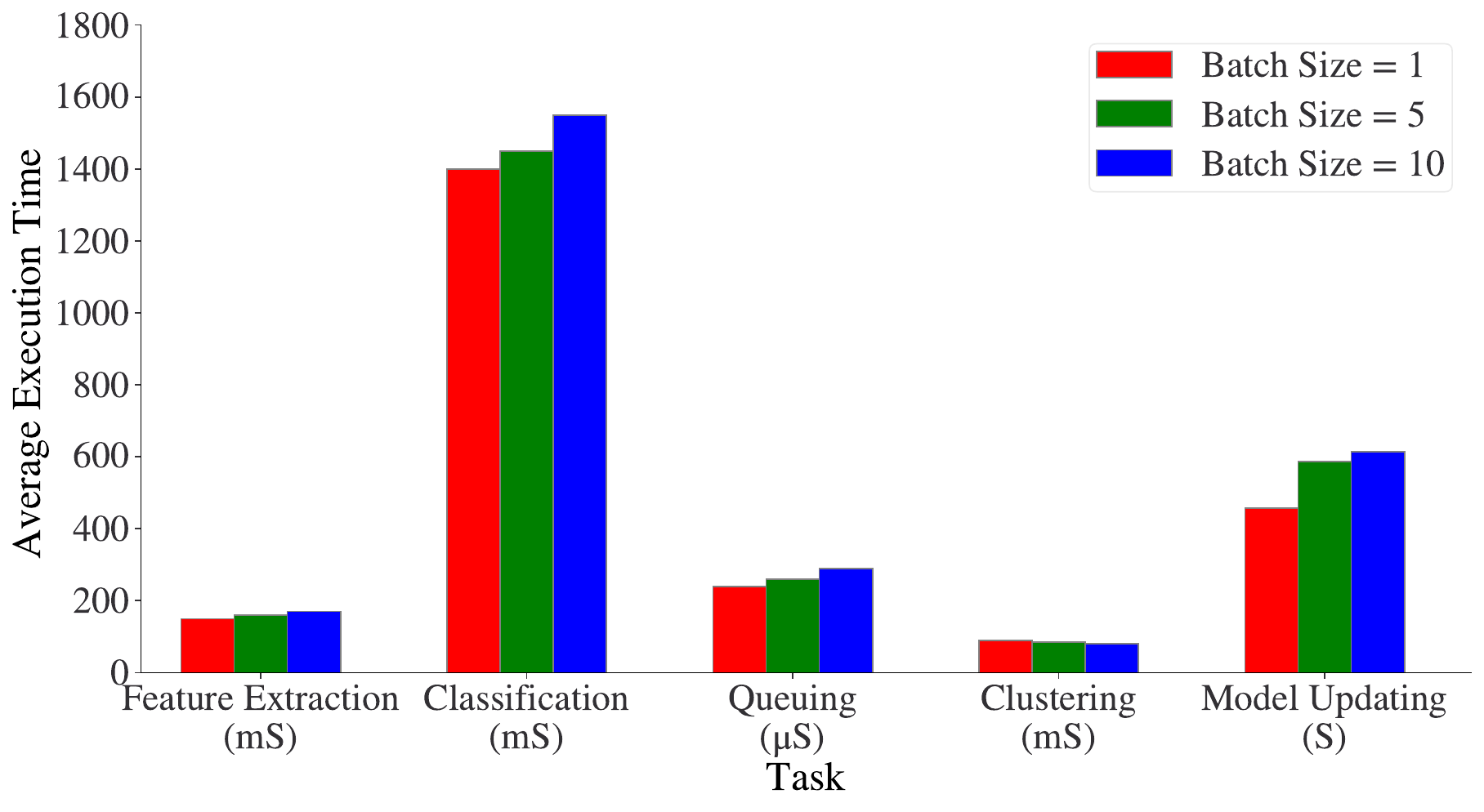}}
\caption{The average execution time of different tasks on Raspberry Pi for different batch sizes}
\label{fig6}
\setlength{\belowcaptionskip}{-50pt}
\end{figure}

\noindent \textbf{\green{OpenSense Execution Time and Resource Consumption on Edge Device:}} We implemented only OpenSense on the Raspberry Pi 4 Model B device since it outperforms the other algorithms in inference, supervised and unsupervised incremental learning, as well as in efficiency. We run the open-world experiment three times for different batch sizes (1,5,10) to simulate periodic data acquisition where each sample in the batch is processed as 1 second-window. In Fig~\ref{fig6}, each task execution time is measured by \{Seconds, mSeconds, \textmu Seconds\} as indicated in the x-axis labels. 

For inference tasks, the latency increases as there are more samples in the batch, but the increase is not significant making the choice of the batch size more flexible depending on system requirement. However, it is important to consider that less samples per batch leads to more inference instances before filling the queue and trigger learning. Although, the execution time of the learning tasks is lower when there is only a sample per batch, it does not imply that choosing lower sample rate is better for the system. The average execution time when clustering the rejected samples is dependent on the queue size regardless of the batch size, while updating the model latency depends on the size of the clusters assigned with a new label. The framework is able to inference and learn new classes from a stream of time-series sensor data.

\green{We also measured the memory usage, CPU utilization, and power consumption of the aforementioned tasks on Raspberry Pi. Table~\ref{tab:effecincy_pi} shows the results. Updating the model is the most energy-consuming task with an average of 3.95W for the training duration. Although some tasks, such as clustering, appear to cause high CPU utilization in this table, they do not have a significant impact on the overall runtime performance of the framework because they have relatively small execution times, as shown in Fig.~\ref{fig6}. Considering the total memory size of Raspberry Pi (2GB), the memory usage of each task is not high when the framework batch size is configured to 10 samples per iteration.  }

\begin{table}[t!]
\strutlongstacks{T}
\centering
\blue{
\footnotesize
  \caption{\green{Memory, CPU and power consumption of OpenSense tasks on Raspberry Pi}}
  \label{tab:effecincy_pi}
    \resizebox{\columnwidth}{!}{\begin{tabular}{|l|c|c|c|}
    \hline
    Task                & Memory Usage        & CPU Utilization & Power Consumption  \\ \hline
    Idle                & -             &  \text{-}   &  2.67W              \\ \hline
    Feature Extraction  & 34.1MB              & 26.8\%      &  3.65W                \\ \hline
    Classification      & 78.3MB              & 27.7\%      &  3.71W                \\ \hline
    Queuing             & 78.9MB              & 30.8\%      &  3.55W                \\ \hline
    Clustering          & 70.1MB              & 79.8\%      &  4.11W                \\ \hline
    Model Updating      & 83.2MB              & 50.8\%      &  3.95W                \\ \hline

    \end{tabular}}
}
\end{table}

\green{}
\begin{figure}[t!]
\centering
\begin{subfigure}{\linewidth}
 \centering
 \includegraphics[width=\linewidth]{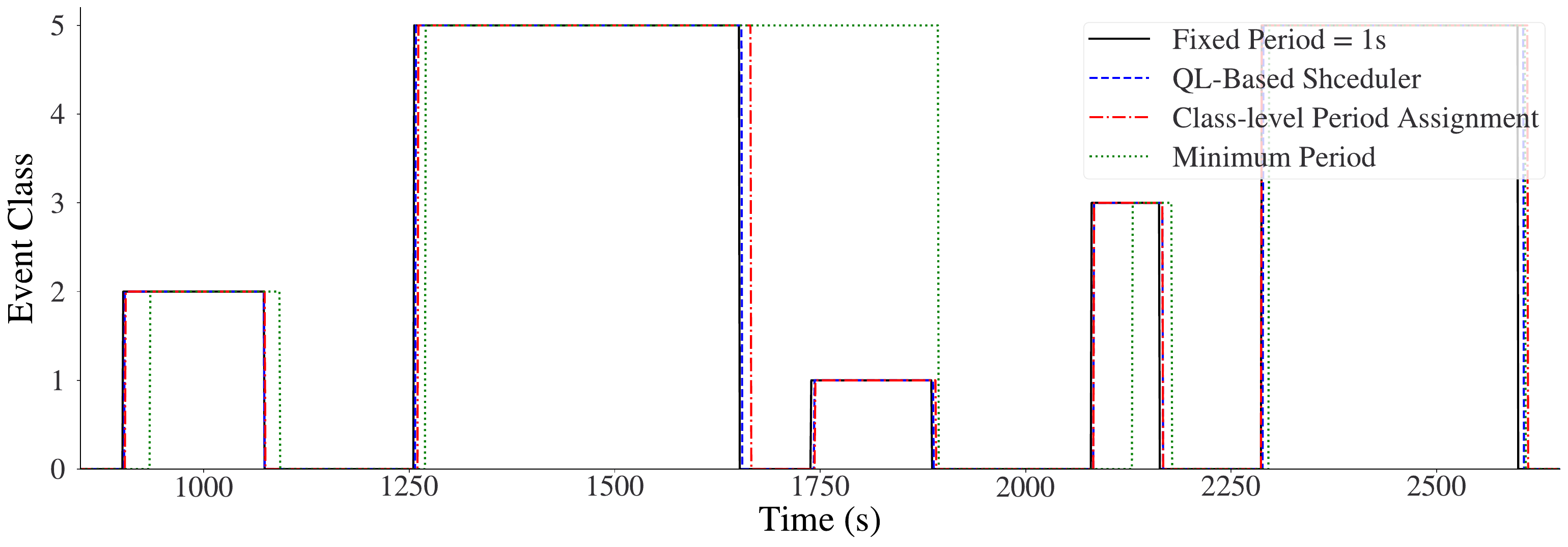}
 \caption{\blue{Sensor data at different periods}}
 \label{fig:sub-first}
\end{subfigure}
\begin{subfigure}{\linewidth}
 \centering
 \includegraphics[width=\linewidth]{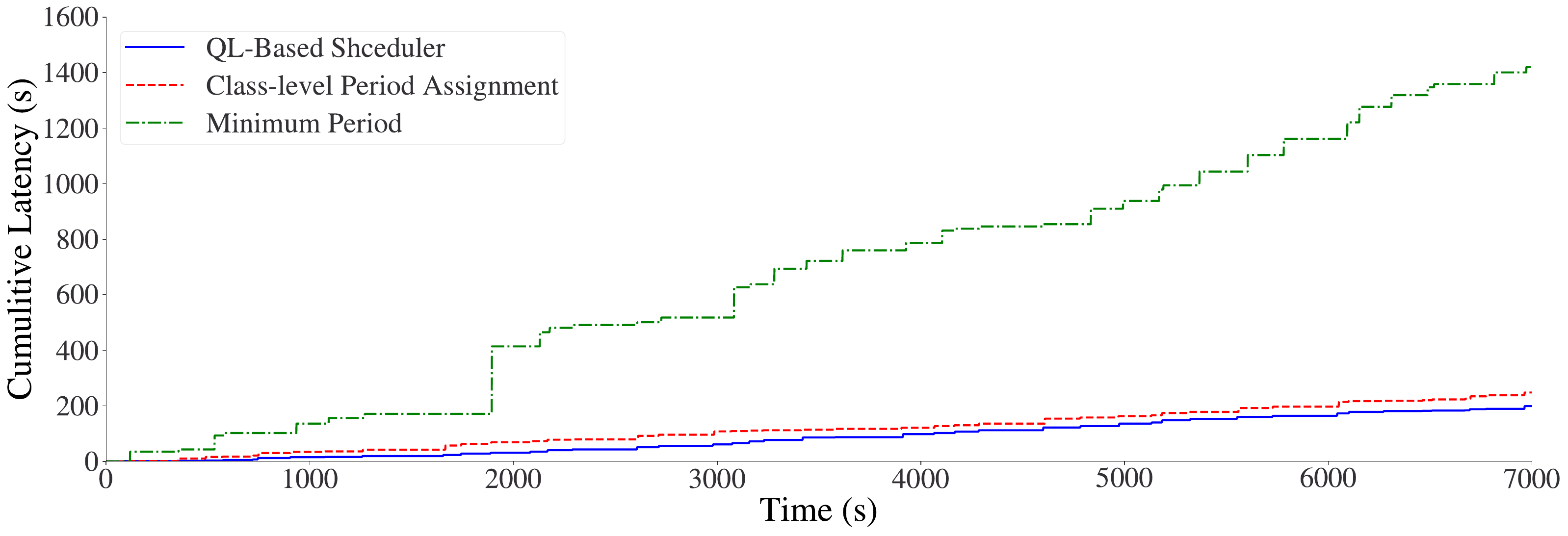}
 \caption{\blue{Cumulative classification latency}}
 \label{fig:sub-second}
\end{subfigure}
\caption{\blue{Classification latency of sensor dynamic schedulers}}
\label{figSD}
\end{figure}

\blue{
\subsection{Results: Sensor Schedulers}
\label{sec:scheuler_results}

\blue{
\noindent \textbf{Classification Latency of Sensor Schedulers:}
We evaluate the classification latency of our proposed sensor schedulers: Class-Level Period Assignment (CLPA) and QL-Based Scheduler (QLBS), in this experiment. 
We compare the proposed schedulers with two other conventional methods: (i) Fixed Period: It uses a fixed period of 1s at all time, resulting in the best classification latency at the cost of excessive energy consumption, and (ii) Minimum Class Interval: It chooses the minimum value among the event durations of each class as the sensing period for that class, thereby choosing longer sensing periods and achieving lower energy consumption than our schedulers but at the cost of longer classification latency. }\blue{Note that for the QLBS, we used the following parameters in the initial training of the model: $n_{episodes}=20,000$, $\epsilon=0.1$, $weight_{p1}=10$ and $weight_{n1}=50$ for $Cr_1$ and $weight_{p2}=1$ and $weight_{n2}=5$ for $Cr_2$.}


\blue{
Fig.~\ref{figSD}(a) illustrates the classification latency patterns over time under the four aforementioned methods. The patterns from both of our schedulers, QLBS and CLPA, closely match that of the fixed 1-sec period pattern, with QLBS being slightly superior to CLPA. As expected, the minimum class interval method not only has the worst classification latency, \green{but also misses the `1: Microwave' event in the middle of the figure.}

Fig.~\ref{figSD}(b) depicts the cumulative latency over time. Overall, both schedulers have significantly lower cumulative latency than the minimum class interval method. Furthermore, QLBS can achieve lower cumulative classification latency compared CLPA, demonstrating the effectiveness of a completely dynamic period selection over class-wise fixed periods. 
}

\blue{
\noindent \textbf{Energy Consumption Reduction by Sensor  Scheduler:} The energy consumption of the sensor board can be significantly reduced with a longer sensing period because the sensor board can enjoy low-power mode during idle time, therefore, require less number of BLE packet transmissions over time. Given that the BLE radio module is in active mode typically dominates the power consumption of small microcontroller-based sensor boards, we consider the normalized number of BLE transmissions compared to a fixed period of 1 second as an energy metric and conduct an experiment based on the simulation results from the previous experiment.

Fig.~\ref{figSE} compares the normalized BLE transmission with respect to the fixed period method under the proposed schedulers and the minimum class interval (lower is better). All three methods including ours achieve a substantial improvement over the fixed period method.
Specifically, the normalized number of transmitted BLE packets under CLPA is approximately 10\% of the total number of transmissions made by the fixed period approach, 18\% for QLBS, and 2\% for the minimum class interval method. While CLPA and QLBS have higher number of BLE transmissions than the minimum class interval method, we conclude that this is an acceptable tradeoff considering the latency improvement achieved by our two proposed schedulers shown in the previous experiment. 
}


%
\begin{figure}[t!]
\centering
 \includegraphics[width=\linewidth]{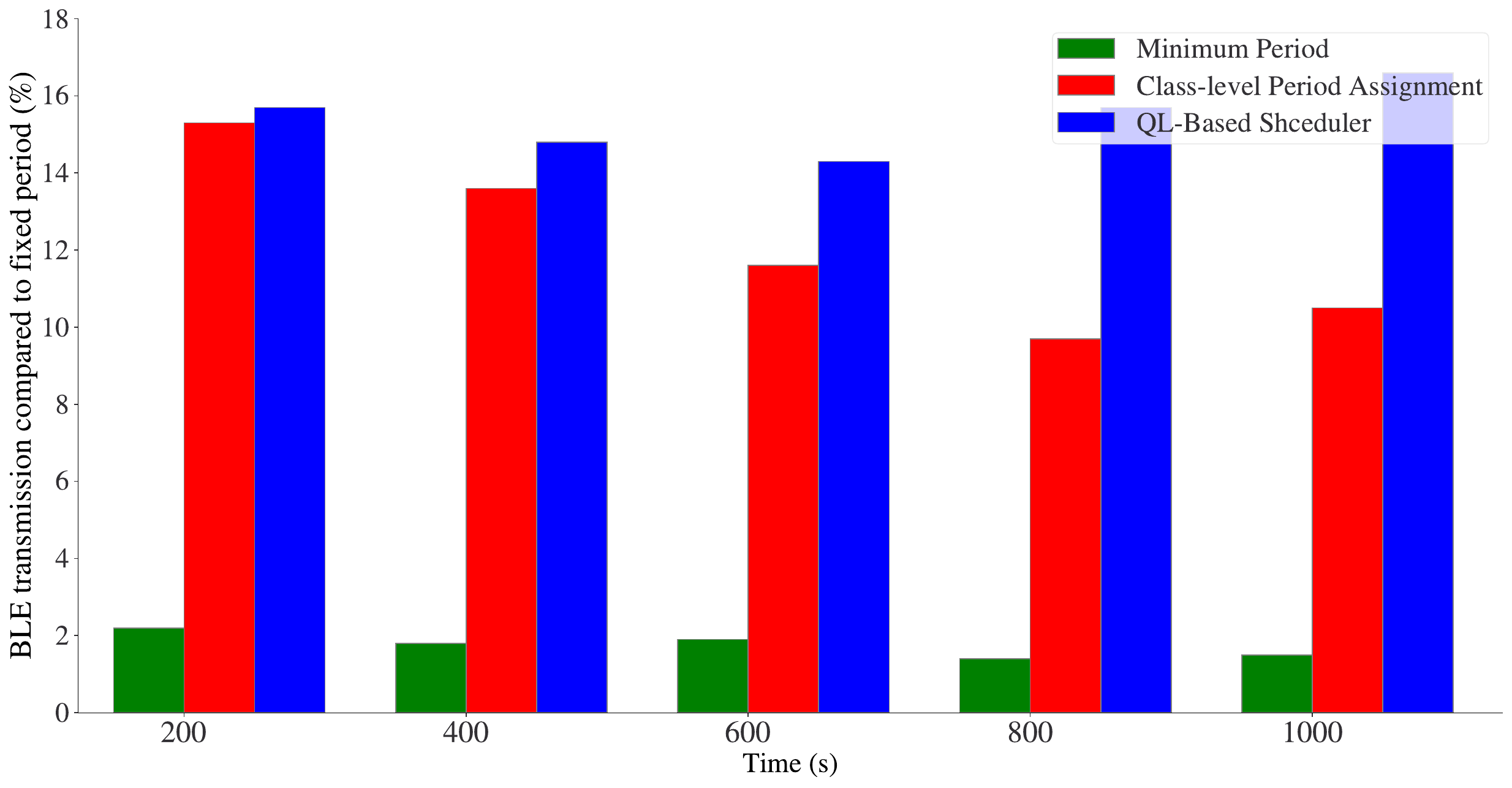} 
 \caption{\blue{Normalized BLE packet transmission w.r.t. the fixed period method (period = 1s)}}
\label{figSE}
\end{figure}

\noindent \textbf{QL-Model Criteria Study:}
In this experiment, we adjust the reward/penalty ratios within the two separate criteria of our QL-based scheduler, $Cr_{1}$ and $Cr_{2}$, to understand their respective influences on the scheduler's performance.  We keep one criterion constant while varying the other, and then alter both simultaneously to capture their individual and combined effects. \blue{ Recall that a reward/penalty is given when ($T_{sp} \:-\:T_{ideal}\:\leq \:CL_s$) or ($T_{sp}\ge prev\_T_{sp}$), corresponding to $Cr_1$ and $Cr_2$ respectively, is met (Alg.~\ref{alg:take_action}).}
Table~\ref{tab:cr12} summarizes the selected combinations that have a notable impact on the scheduler's performance. 
If we increase the reward and decrease the penalty in $Cr_{1}$, e.g., from 1st row to 2nd row, the scheduler likely gives longer sensing periods $T_{sp}$, thereby reducing BLE packing transmissions; however, this adversely affects the classification latency performance (more $CL_s$ misses). Similarly, if we increase the reward in $Cr_{2}$, the sensing periods further increase, as can be seen from the 5th and 6th rows. \blue{Conversely, we observe that increasing the reward of $Cr_{1}$ and lowering it for $Cr_{2}$, provides an acceptable tradeoff between higher BLE transmissions and lower $CL_s$ missing rate, e.g., 3rd and 4th rows.}
}

\begin{table}[t!]
\centering
\blue{
\footnotesize
  \caption{\blue{Q-learning model with different $Cr_{1}$ and $Cr_{2}$ reward/penalty (R/P) ratios}}
  \label{tab:cr12}
  \resizebox{\columnwidth}{!}{\begin{tabular}{|c|c|c|c|}
    \hline
    $Cr_{1}$ (R/P) & $Cr_{2}$ (R/P) & \makecell{Norm. BLE Trans.} & $CL_s$ Misses\\ \hline
    10/50 & 5/1 & 19\% & 0 \\ \hline
    50/10 & 5/1 & 12\% & 7 \\ \hline
    50/50 & 5/1 & 12\% & 1 \\ \hline
    50/10 & 10/1 & 13\% & 2 \\ \hline
    50/10 & 1/10 & 6\% & 20 \\ \hline
    50/10 & 50/10 & 4\% & 34 \\ \hline
    50/50 & 20/20 & 8\% & 13 \\ \hline
    20/20 & 50/50 & 1\% & 34 \\ \hline
  \end{tabular}}
}
\end{table}
\blue{
\noindent \textbf{Sensor Scheduling Overhead on Edge Device:} We analyze the execution time of our sensor scheduling algorithms to make a ``single scheduling decision'' on an edge device. The single scheduling decision is defined here as the operations performed by each algorithm to find one $T_{sp}$ value and evaluate whether it meets the $CL_s$ constraint or not. In QLBS, this involves determining the best action to take from a trained Q-table given the current state and assessing a reward or penalty based on this action. As for CLPA, a scheduling decision is simply made by selecting the $T_{sp}$ that has been found for the current event and by comparing it to the $CL_s$ constraint. Obviously, CLPA requires a much shorter execution time compared QLBS. \green{We measure the execution time (mean, min, max, stddev) as well as the memory usage, CPU utilization, and power consumption} of our schedulers on a representative embedded edge device, Raspberry Pi 4, and report the results in Table~\ref{tab:overhead}. The average execution time of QLBS is 115$\mu$s, whereas CLPA averages only 6.4$\mu$s. 
 Although the difference can be seen large, this does not impose a significant overhead on the edge device in practice because it is executed only once per sensing period which typically spans several seconds. \green{Additionally, memory usage, CPU utilization, and power consumption are all relatively low for both scheduling algorithms.}
}

\begin{table}[t!]
\centering
\blue{
\footnotesize
  \caption{\green{Sensor scheduler overhead on Raspberry Pi}}
  \label{tab:overhead}
  \resizebox{\columnwidth}{!}{\begin{tabular}{|c|c|c|c|c|c|c|c|}
    \hline
    Algorithm & Mean  & Min & Max & StdDev & Mem. & CPU & Pwr. \\ \hline
    CLPA & 6.4$\mu$s & 5.2$\mu$s & 45$\mu$s & 3$\mu$s  & 14.4MB  & 6.4\%  &  3.08W \\ \hline
   QLBS & 115$\mu$s & 35$\mu$s & 412$\mu$s & 73$\mu$s  & 18.3MB  & 7.1\%  &  3.28W \\ \hline

  \end{tabular}}
}
\end{table}

\begin{table}[t!]
\centering
\blue{
\footnotesize
  \caption{\blue{Updating sensor scheduler on Raspberry Pi}}
  \label{tab:scheduler_update}
    \resizebox{\columnwidth}{!}{\begin{tabular}{|l|l|l|l|}
    \hline
    Sensor Sched. Alg.   & Sched. Update Time & Avg. $CL_s$ Misses & Norm. BLE Trans.             \\  \hline
    QLBS with $\theta$=0.1    & 5.6                     & 29.4            & 6.1\%                          \\ \hline
    QLBS with $\theta$=0.01   & 31.5                    & 14.7            & 8.2\%                          \\ \hline
    QLBS with $\theta$=0.001  & 340.9                   & 9.6             & 10.4\%                         \\ \hline
    QLBS with $\theta$=0.0001 & 2364.9                  & 5.2             & 10.7\% \\ \hline
    CLPA                 & 0.01            & 12.8        & 10\%       \\     \hline               
    \end{tabular}}
}
\end{table}

\blue{
\noindent \textbf{Updating Sensor Schedulers on Edge Device:} 
In this experiment, we run both sensor scheduling algorithms, CLPA and QLBS, on the edge device to find the time it takes for Alg.~\ref{alg:class_level_sched} to update the $T_{sp}$ for each class, and for Alg.~\ref{alg:train_QL} to update the Q-table, respectively. The schedulers are updated once there are sufficient new time intervals $T_e$ for each class. To compare both algorithms fairly, we use the same sensing history data for both. 
In this experiment, $\theta$ threshold for QLBS is varied from 0.1 to 0.0001 to assess the tradeoff between updating time and output quality, while the $n_{success}$ parameter remains fixed at 5. 
In Table ~\ref{tab:scheduler_update}, we show the time needed to update each algorithm and the corresponding performance metrics, e.g., $CL_s$ misses and BLE packet transmissions (lower is better for both), achieved from each update. We see that CLBA requires only 0.01 second to update its $T_{sp}$ values, which is substantially small compared to an average of 5.6 to 2364.9 seconds to update QLBS for different thresholds values. \green{When updating QLBS once with each $\theta$ value in Table~\ref{tab:scheduler_update}, the average memory usage we observed is 27.2MB. 
} 
Although CLPA can update in a very short time, the outcome of the QLBS update can be more beneficial to the edge device in terms of fewer $CL_s$ misses and lower BLE packet transmissions, and the tradeoff can be controlled with the $\theta$ parameter.} 

\blue{
\noindent \textbf{Sensor Scheduler Performance on PAMAP2 Dataset:} To assess the applicability of the proposed algorithms to other types of applications, we train and evaluate CLPA and QLBS on the PAMAP2 dataset. In Table ~\ref{tab:scheduler_pamap2}, QLBS achieves a lower $CL_s$ miss rate and simultaneously yields a much higher BLE packet transmissions than CLPA. \blue{The reason behind this is due to the fact that some events in the PAMAP2 dataset exhibit rather sporadic and extended durations, e.g., a single continuous walking session is 800s, which is more favorable to CLPA since it allows it to find longer sensing period $T_{sp}$ for that class. We conclude that both algorithms can adapt to different datasets but QLBS can be more advantageous than CLPA in dynamic environments.}}

\begin{table}[t!]
\footnotesize
\centering
\blue{
  \caption{\blue{Evaluating sensor scheduler on PAMAP2 dataset}}
  \label{tab:scheduler_pamap2}
    \begin{tabular}{|l|l|l|}
    \hline
    Sensor Scheduler Alg. & Avg. CL Misses & Norm. BLE Trans. \\ \hline
    CLPA                       & 19              & 4.6\%              \\ \hline
    QLBS                       & 12              & 16.7\%        \\ \hline    
    \end{tabular}
}
\end{table}

\begin{figure}[t!]
 \includegraphics[width=\linewidth]{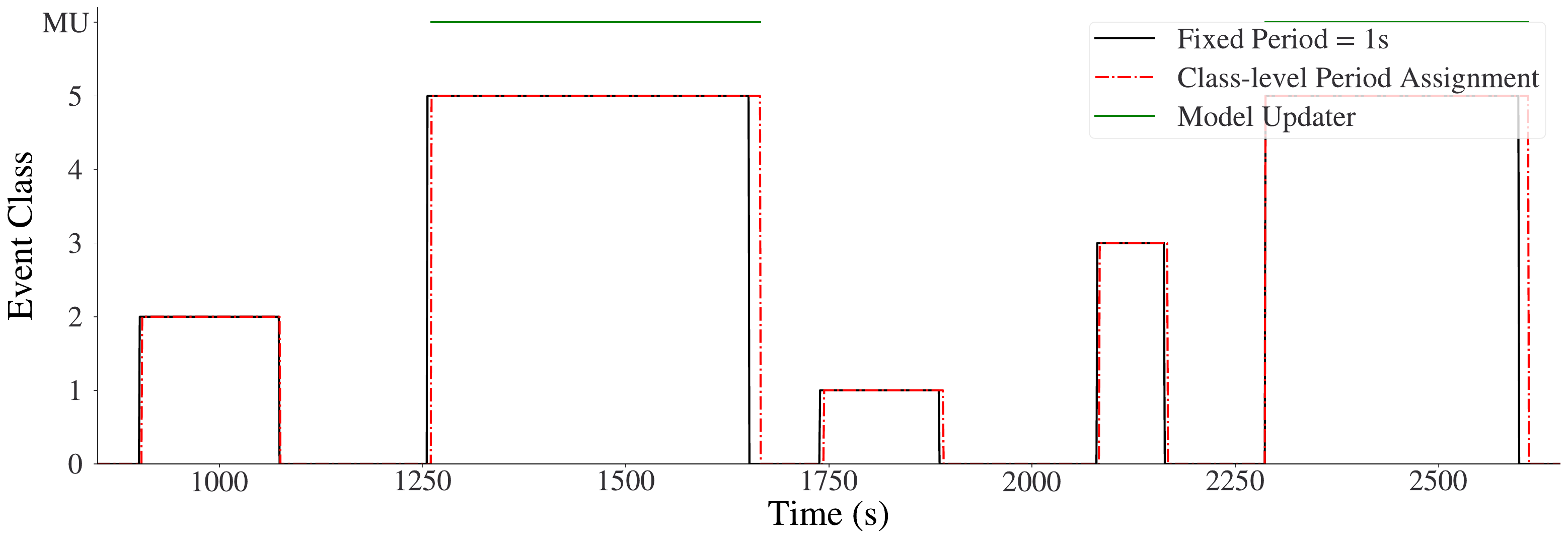} 
 \caption{\blue{Case study of our model updater for vent fan events}}
\label{figMU}
\setlength{\belowcaptionskip}{-50pt}
\end{figure}

\blue{
\noindent \textbf{Case Study for Model Updater:} This experiment is to demonstrate the functionality of our updater in the context of updating the inference model ($update\_mode=classifier$). It builds upon the previous experiment, ``Classification Latency of Sensor Schedulers'', and introduces a scenario where 100 samples of an unknown class that has been detected through the online clustering algorithm. We choose to use CLPA here since it assigns a period of 33s to the vent fan event that satisfies the least amount of time to update the model with 1 sample (31s). In Fig~\ref{figMU}, the green lines indicate that the updater is triggered during the vent fan event and successfully updates the model with all the samples during the testing period. 
}

\section{Conclusion}
\blue {
This paper presents OpenSense, a framework that is capable of discovering and learning new classes incrementally from a stream of time-series sensing data, and scheduling sensors and model updates on edge devices. The EVM model adopted in our framework is shown to be effective in rejecting unknown samples correctly while being computationally efficient during incremental learning compared to other algorithms. Our proposed sDNN architecture not only outperforms the state-of-an-art, but also converges faster. Furthermore, the proposed sensor schedulers allow the framework to be efficiently idle without compromising the inference latency, as well as enabling the model to be updated at runtime on the edge device. With these features, our framework can be successfully deployed on a resource-constrained edge device for various applications. 
This framework opens a broad range of interesting research directions. One direction can be utilizing heterogeneous accelerators and GPUs in edge devices to minimize the computational overhead with proper energy management support. It would be also interesting to extend this framework to work with intermittent batteryless energy-harvesting devices, which could be extremely challenging due to the limited memory and energy resources available on these devices.}


\bibliographystyle{IEEEtran}
\bibliography{ref.bib}

\vspace{12pt}

\end{document}